\documentclass[a4paper,12pt]{article}
\usepackage{mathptmx}
\usepackage{epsfig}
\usepackage{float}
\usepackage{amssymb}
\usepackage{latexsym}
\usepackage{authblk}
\usepackage{chngcntr}
\usepackage{amsfonts}
\newcommand{\beq}{\begin{equation}}
\newcommand{\eeq}{\end{equation}}
\newcommand{\beqa}{\begin{eqnarray}}
\newcommand{\eeqa}{\end{eqnarray}}

\usepackage{natbib}
\usepackage{graphicx}
\usepackage{multirow}
\begin{document}



\title{Ionization effect in the Earth's atmosphere during the sequence of October-November 2003 Halloween GLE events}
\author[1,2]{A.L. Mishev}
\author[3]{P.I.Y. Velinov}
\affil[1]{Space Physics and Astronomy Research Unit, University of Oulu, Finland.}
\affil[2]{Sodankyl\"a Geophysical Observatory, University of Oulu, Finland.}
\affil[3]{Institute for Space Research and Technology, Bulgarian Academy of Sciences, Sofia, Bulgaria}
\maketitle
\begin{abstract}
The effect of precipitating high-energy particles on atmospheric physics and chemistry is extensively studied over the last decade. In majority of the existing models, the precipitating particles induced ionization plays an essential role. For such effects, it is necessary to possess enhanced increase in ion production, specifically during the winter period. In this study, we focus on highly penetrating particles -  cosmic rays. The galactic cosmic rays are the main source of ionization in the Earth's stratosphere and troposphere. On the other hand, the atmospheric ionization may be significantly enhanced during strong solar energetic particle events, mainly over the polar caps. A specific interest is paid to the most energetic solar proton events leading to counting rate enhancement of ground-based detectors, namely the so-called ground level enhancements (GLEs). During solar cycle 23, several strong ground level enhancements were observed. A sequence of three GLEs was observed in October-November 2003, the Halloween events. Here, on the basis of 3-D Monte Carlo model, we computed the energetic particles induced atmospheric ionization, explicitly considering the contribution of cosmic rays with galactic and solar origin. The ion production rates were computed as a function of the altitude above sea level using reconstructed solar energetic particles spectra. The 24 hours and event averaged ionization effects relative to the average due to galactic cosmic rays were also computed.
\end{abstract}

 \normalsize

\label{cor}{\small For contact: alexander.mishev@oulu.fi}


\section{Introduction}
Various populations of precipitating high-energy particles and/or high-energy radiation contribute to the atmospheric ionization viz. subatomic particles - cosmic rays of galactic or solar origin, precipitating particles from radiation belts, auroral electrons, UV and X-ray solar emissions \citep[for details see the reviews by][as well as the references therein]{Vainio2009, Mironova2015}. Energetic particles induced ionization in the upper part of the Earth's atmosphere is dominated by solar UV and X-rays. At altitudes below about 100 km above sea level (a.s.l.), specifically in the stratosphere and upper troposphere, the atmospheric ionization is essentially due to the penetrating flux of galactic cosmic rays (GCRs), which is variable to a small degree \citep[for details see][]{Potgieter2013}, and thus leading to quasi-constant ion production background \citep[for details see][]{Bazilevskaya08}. In some cases, as a result of solar eruptions, an enhanced high-energy particles flux of solar origin, known as solar energetic particles (SEPs), can impinge the Earth atmosphere, eventually leading to significant ion production \citep[e.g.][]{Uso11b}. 

The energetic particles precipitation induced ionization in the troposphere and stratosphere results from the nuclear-electromagnetic-lepton cascade, i.e. sequence of successive interactions of the penetrating primary cosmic ray particle with the atmospheric particulates, producing a large amount of secondary particles losing their energy mainly via ionization \citep[e.g.][]{Dorman04, Greider11, Tanabashi2018}. The maximum of ion production in the atmosphere due to cosmic rays (CRs) is observed at the altitude of about 12-15 km a.s.l., known as Regener-Pfotzer maximum \citep{Regener1935}. The GCR flux is slightly modulated in the Heliosphere and inversely follows the 11-year solar cycle, responding also to transient phenomena leading to short term episodes as the Forbush decreases \citep{Forbush37, Forbush58}. In some cases, the sporadically occurring SEPs possess energy of about a 1 GeV/nucleon or even greater, as a result induce similarly to GCRs atmospheric shower. Hence, in that case, secondaries produced by SEPs penetrate deep into the atmosphere or even reach the ground leading to ground level enhancements (GLEs) \citep[e.g.][]{Poluianov2017}. Therefore, GLEs may cause a significant excess of ionization, specifically over the polar caps \citep[e.g.][]{Jackman200011659, Jackman20116153, Mis11a, Uso11b, Mishev201578, Mitthumsiri20177946, Mishev2018316}.

A systematic study of the induced by high-energy particles impact ionization allows one to clarify the influence of precipitating particles on different atmospheric chemistry processes, global electric circuit and atmospheric physics, specifically on minor components \citep[e.g.][]{Krivolutsky20061602, Randall2007, Jackman2008765, Rozanov2012483,  Nicoll2014, Verronen2015381, Sinnhuber20181115}. Naturally, over the last decade, the energetic particles impact ionization was extensively studied. Thus, the increased high-energy particles flux during GLEs provide a unique opportunity to study such effects in enhanced mode. During the solar cycle 23, sixteen GLEs were observed, the first event occurred on 6 November 1997 (GLE $\#$ 55),  the last occurred on 13 December 2006 (GLE $\#$ 70), the full list is available at the Oulu Cosmic Ray Station  \texttt{http://gle.oulu.fi}) \citep[e.g.][]{Gopalswamy201223}.  GLEs occur sporadically,  differ from each other in spectra, particle fluence, anisotropy, duration, apparent source position, geomagnetic conditions \citep[e.g.][]{Moraal201285}. Therefore, they are usually studied case by case. Here, we focus on the sequence of three consecutive GLEs, viz. the three so-called Halloween events of October-November 2003, occurred on 28 October (GLE $\#$ 65), 29 October (GLE $\#$ 66), and on 2 November (GLE $\#$ 67), respectively. On the basis of precisely derived SEP spectra and convenient state of the art model, we assessed the ion production and the corresponding ionization effect during the three Halloween GLE events. 

\section{Employed model for computation of induced ionization}
The ion production in the atmosphere due to precipitating high-energy particles can be assessed by analytical (parametrization) and/or semi-empirical models \citep[e.g.][]{Bri70, Vit96}. However,  parameterization models usually exhibit constraints to a given atmospheric region, atmospheric cascade component or primary particle \citep[e.g.][and references therein]{Velinov2013}. On the other hand, models based on Monte Carlo simulations of the atmospheric cascade allow one to reliably assess the ion production rate considering all the physical processes \citep{Desorgher05, Usos06, Velinov2009, Paschalis201426, Banjac201950}. Here, we employed a model similar to \citet{Usos06}, with the full description and application are given elsewhere \citep[i.e.][]{Mishev2007225, Velinov2009, Mishev2015360}. The ion production rate as a function of the altitude a.s.l. is given by:

\begin{equation}
  q(h,E) =  \frac{1} E_{ion} \sum_{i} \int_{E_{cut}(R_{c})}^{\infty} \int_{\Omega} D_{i}(E) \frac{\partial E(h,E)}{\partial h} \rho(h)dE d\Omega 
        \label{simp_eqn1}
   \end{equation}

where $\partial  E$ is the deposited energy in an atmospheric layer $\partial  h$ , $h$ is the air overburden (air mass) above a given altitude in the
atmosphere expressed in $g/cm^{2}$ or altitude a.s.l., $D_{i}(E)$ is the differential cosmic ray spectrum for a given
component $i$: protons p, Helium ($\alpha$-particles), the latter representative for heavier nuclei with atomic number Z $>$ 2 \citep[for details see][]{Usos06, Mis11}, $\rho$ is the atmospheric density in $g.cm^{-3}$, $E$ is the initial energy of the incoming primary nuclei on the top of the atmosphere, $\Omega$ is a solid angle and $E_{ion}$ = 35 eV is the average energy necessary for creation of an ion pair in air \citep{Porter1976154}. The integration is over the kinetic energy $E_{cut}(R_{c})$ above the rigidity cut-off $R_{c}$ for a nuclei of type $i$ at a given geographic location by the expression:

\begin{equation}
 E_{cut,i}=\sqrt{ \left( \frac {Z_{i}} {A_{i}}\right)^{2} R_{c}^{2}+ E_{0}^{2}} - E_{0}
        \label{simp_eqn2}
   \end{equation}
  
\noindent where $E_{0}$ =  0.938 GeV/n is the proton's rest mass.

Here, the computation of the cut-off rigidity was performed with the MAGNETOCOSMICS code, explicitly considering the measured K$_{p}$ index during the events \citep{Desorgher05}, assuming a combination of the IGRF geomagnetic model as the internal field model and the Tsyganenko 89 model as the external field \citep{Tsyganenko89}. This combination allows one to compute straightforwardly with reasonable precision the local rigidity cut-offs over the globe \citep{Kudela04, Kud08, Nevalainen13}. We emphasize that employment of newer version of Tsyganenko models e.g. Tsyganenko 96 or Tsyganenko 01 \citep{Tsyganenko95, Tsyganenko2002}, would lead to comparable results since the other model uncertainties such as ionization yield function and the used as input SEP spectra are considerably greater as discussed below.

During GLEs, ion production in the atmosphere is presented as a superposition of GCRs and SEPs contributions. For the contribution of GCRs in Eq. (1) the force field model is employed \citep{Gle68, Bur00}, where the parametrization of the local interstellar spectrum is according to \citet{UsoskinJGR2017} and the modulation potential is taken from \citet{Usoskin11a}. Accordingly, the SEPs spectra in equation (1) are considered according to \citet{Miroshnichenko2005, Kocharov2017}. 

Note, that for the computations a realistic atmospheric model NRLMSISE 00 is employed, that explicitly considers the specifics of the atmospheric conditions and type of the incident primary particle \citep{Picone2002, Mishev2010476, Mishev14d}.

\section{Sequence of three Halloween GLEs on October-November 2003}
A violent solar activity was observed in October--November 2003, which produced a sequence of three consecutive GLEs, with onsets occurring on 28 October, 29 October and on 2 November, respectively \citep{Gopalswamy2005, Liu20061135, Gopalswamy201223}. The events were observed by the global neutron monitor network, the details and count rate records are given in great detail in \texttt{http://gle.oulu.fi}). 

The GLE  $\#$ 65 on 28 October 2003 was associated with a large flare (4B, X17.2) occurred in the active region AR10486. The GLE $\#$ 65 followed significant interplanetary disturbance related to previously ejected coronal mass ejection (CME) on 26 October with correspondence with a 3B/X1.2 flare in the same active region. The SEP spectra were relatively hard with observed softening throughout the event (Fig.1a). 

The GLE  $\#$ 66, occurred on 29 October, was characterized by a notably smaller recorded neutron monitor count rate increases, thus this event was considerably weaker. Softer SEP spectra were observed, with non-significant variation throughout the event (Fig.1b). A strong Forbush decrease was also observed prior to and during this event, which was explicitly considered during our computations of ion production rate, i.e. a GCR flux reduction was taken into account during the computations. 

The GLE  $\#$ 67, occurred on 2 November 2003, was related to an X8.3/2B solar flare, with onset at about 17:30--17:35 UT. In general, the event was characterized by a large anisotropy in its initial phase and relatively hard but constantly softened SEP spectra (Fig.1c), the details are given in \citep{Kocharov2017}.

It should be noted that in  this period occurred the strongest in 30 years (since march 1989) geomagnetic storm. The planetary $A_{p}$ index reached values $A_{p}$=204 on 29 October 2003, $A_{p}$=191 on 30 October 2003 and $A_{p}$=116 on 31 October 2003.

\begin{figure}
\centering
\includegraphics[width=0.8\textwidth]{{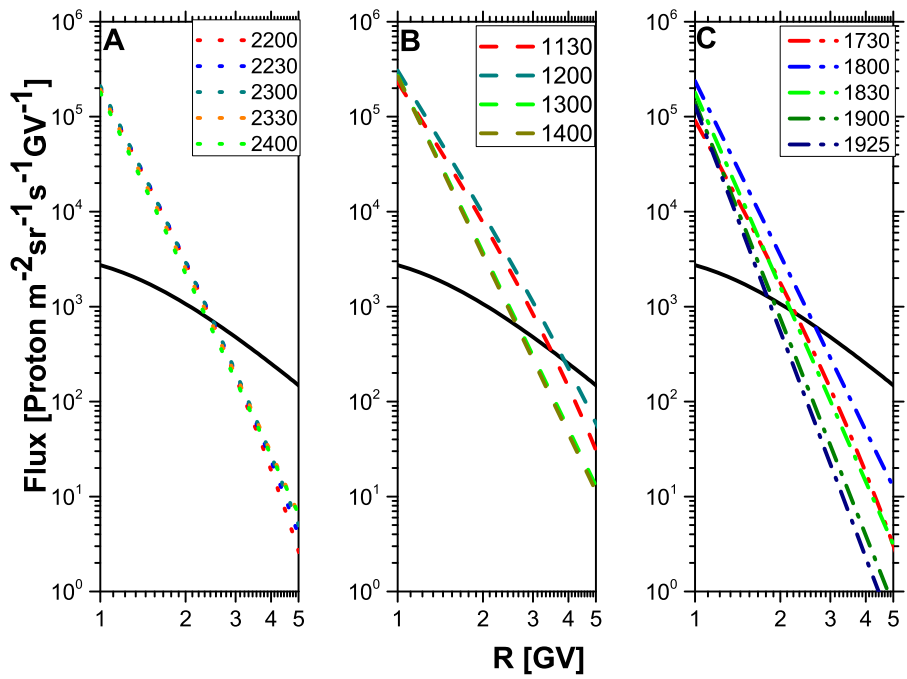}}
\caption{Rigidity spectra of SEPs during selected stages of Halloween GLEs as denoted in the legend. Panels a,b,c, correspond to  GLE $\#$ 65 occurred on 28 October 2003, GLE $\#$ 66 occurred on 29 October 2003 and GLE $\#$ 67 occurred on 2 November 2003, respectively. The solid line denotes the GCR flux. }. 
\label{Fig1}
\end{figure}  

The presented in Fig.1 SEP spectra were used as input in Eq. (1) for the computation of ion production rate throughout the events. The spectra were mostly derived on the basis of a verified method for data analysis of ground-based neutron monitor data \citep[e.g.][]{Mishev2018sol72, Mishev16SF}, which reliably accounts for the high-energy part of the spectra, usually underestimated by space-borne instruments \citep[for details see][]{koldobskySF19}. The derived GLE particles spectra are with good agreement with \citet{koldobskySF19}. More information about the used spectra is presented in \citet{Miroshnichenko2005, Kocharov2017}. Here, we would like to point out that the input SEP spectra in the model for computation of the ion production is crucial and can lead to an important uncertainty, up to an order of magnitude, in the computations similarly to \citet{Butikofer2015}. The other model uncertainties, namely the Monte Carlo simulations of the atmospheric cascade used for the computation of the ionization yield function \citep[e.g.][]{Mishev2010476}, integration methods in Eq. (1) and computation of the cut-off rigidities using the  corresponding magnetospheric model are significantly smaller. Therefore, the employment of SEP spectra based on a verified method is specifically important.    

\section{Atmospheric ionization during Halloween GLEs}
The Halloween GLEs differ in duration, features, accordingly anisotropy and spectra, which play an essential role for the computations of ion production throughout the events \citep{Moraal201285}. In addition, GLE $\#$ 66 occurred during deep Forbush decrease, which was explicitly considered during our computations, i.e. the GCR flux was adjusted from NMs measurements, the data retrieved from neutron monitor database \texttt{www.NMDB.eu} \citep{Mav11}. Accordingly, the GLE $\#$ 67 occurred during the recovery phase of a Forbush decrease, which was taken into account in a similar way.

\subsection{Ion production rates during Halloween GLEs}
Using the model described in Section 2 and the derived GLE particles spectra (Section 3) we computed the ion production rate in the stratosphere and troposphere from ground level to about 35 km a.s.l. considering the GCRs and SEPs contribution \citep{Mishev2018CR, Mishev2020CR}. In Fig.2 we present the ion production during the Halloween GLEs, specifically in the polar and sub-polar region with rigidity cut-off $R_{c}$ $\le$ 1 GV (Fig.2a,b,c) and high mid-latitudes region with rigidity cut-off $R_{c}$ $\approx$ 2 GV (Fig.2d,e,f).

The computed during GLE $\#$ 65 ion production rate was significant throughout the initial and main phase of the event, specifically in the polar low stratosphere (Fig. 2a). The ion production rate diminished but remained significant during the late phase of the event. However, in a region with rigidity cut-off of about $R_{c}$ $\approx$ 2 GV the ion production rate was comparable to the average due to GCRs (Fig.2d). Moreover, at altitudes of about 10 km a.s.l. and below the ion production rate due to GCR was greater than that due to SEPs, because of the soft spectra of the latter. Accordingly, in the region of $R_{c}$ $\approx$ 3 GV,  the ion production due to GCR dominated in the whole atmosphere. The time evolution of the ion production rate at an altitude of 15 km a.s.l. during  GLE $\#$ 65 is presented in the Appendix (Fig.A.1).
 
The computed ion production rate during GLE $\#$ 66 was considerably lower (Fig.2b), because of softer SEP spectra compared to GLE $\#$ 65, the notably reduced SEP flux and the accompanying deep Forbush decrease. The ion production rate during GLE $\#$ 66 was slightly variable throughout the event with a tendency for diminishing. At polar and sub-polar region with rigidity cut-off $R_{c}$ $\le$ 1 GV, the ion production rate was still significant, however, in region with rigidity cut-off $R_{c}$ $\approx$ 2 GV the contribution of SEPs was smaller than that due to GCRs (Fig.2e). Similarly to GLE $\#$ 65, the time evolution of the ion production rate at an altitude of 15 km a.s.l. during  GLE $\#$ 66 is presented in the supplementary material Appendix (Fig.A.2).

The ion production rate during the last event - GLE $\#$ 67 was greater than GLE $\#$ 66, specifically during the initial and main phase of the event, but rapidly diminished during the late phase (Fig.2c). In addition, because the softer SEP spectra compared to GLE $\#$ 65 the computed ion production rates were slightly below that the computed during the GLE $\#$ 65. In the region of mid-latitudes with $R_{c}$ $\approx$ 2 GV, the contribution of SEPs was smaller than that due to  GCRs (Fig.2f). The time evolution of the ion production rate at an altitude of 15 km a.s.l. during  GLE $\#$ 67 is presented in the Appendix (Fig.A.3).

\begin{figure}
\centering
\includegraphics[width=0.8\textwidth]{{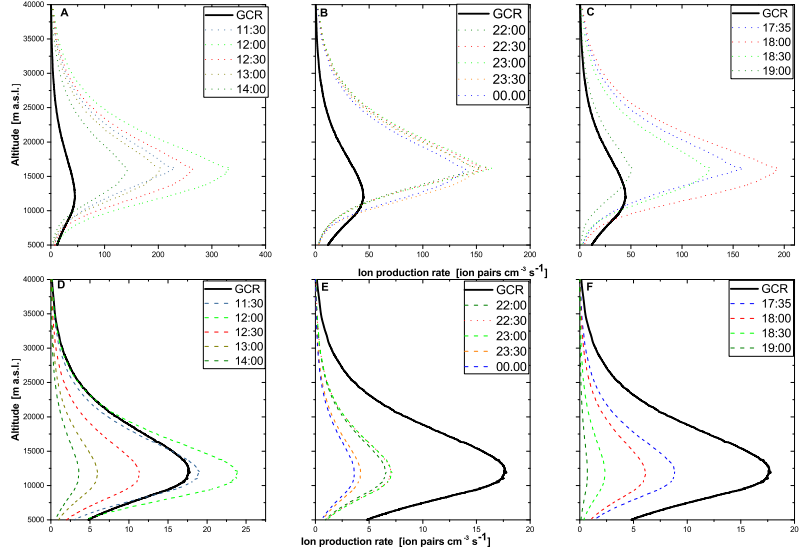}}
\caption{Ion production rates due to CRs during Halloween GLEs on October-November 2003. Panels a,b,c, and d,e,f, correspond to GLE $\#$ 65, 66, 67 respectively. The top panels correspond to region with  $R_{c}$ $\le$ 1 GV, the bottom panels correspond to region with  $R_{c}$ $\approx$  2 GV.}. 
\label{Fig2}
\end{figure}  

One can see that the computed ion production rates were greatly variable throughout the events. The ion production was significant in the polar region and considerably diminished in region with $R_{c}$ $\approx$ 2 GV. The ion production rates were maximal during the strongest event - GLE $\#$ 65. In all cases, the contribution to the ion production due to GCRs in the polar region was smaller that that due to SEPs, but dominated at region with $R_{c}$ $\approx$ 2--3 GV. 

\subsection{Integrated ionization effect in the atmosphere}
In order to compute the ionization effect during GLEs, specifically for atmospheric physics and chemistry purposes, it is more convenient to perform integration over selected time scale(s), corresponding to event duration and/or 24 hours. Here, using the computed ion production rates during the Halloween GLEs, we computed the corresponding ionization effect. The ionization effect represents the averaged over the event or 24 hours ion production rate during a GLE considering the SEP and actual GCR contributions versus averaged ion production rate due to GCR prior to the event, assuming recombination model similarly to \citet{Krivolutsky20061602}.

The results of those computations for 24 hours averaged ionization effect are shown in Figs.3--5 for GLE $\#$ 65, GLE $\#$ 66 and GLE $\#$ 67, respectively. Note, that here we computed the cut-off rigidity during the events, explicitly considering the complex goemagnetospheric conditions employing a combination of IGRF \citep{Thebault2015} and Tsyganenko 89 \citep{Tsyganenko89} models, as discussed above. This allowed us to perform realistic high-precision computations of the global distribution of ion production over the globe, accordingly the corresponding ionization effect. 

\begin{figure}
\centering
\includegraphics[width=0.8\textwidth]{{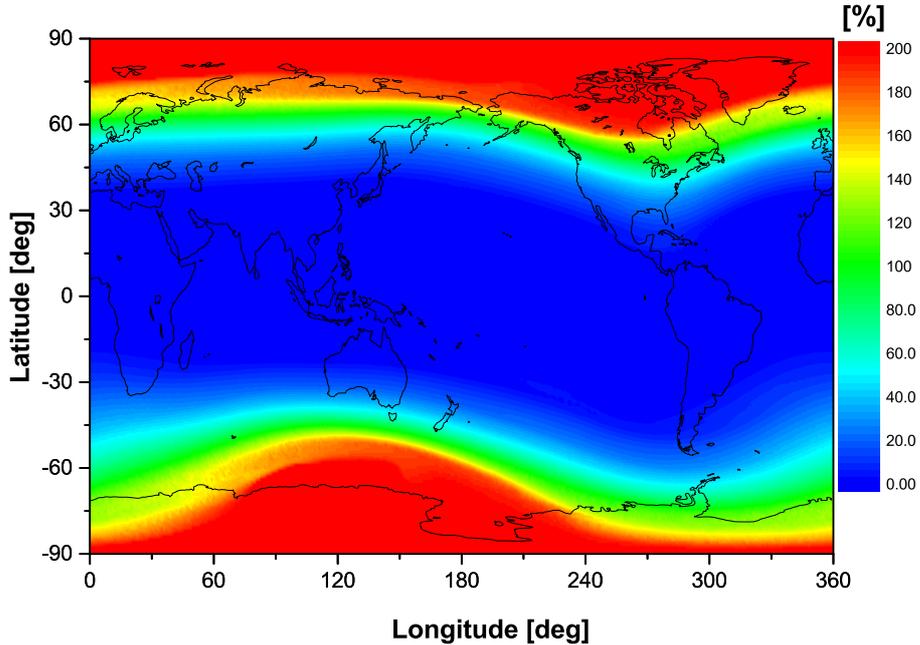}}
\caption{Map of the 24$^{h}$ averaged ionization effect in the region of Regener-Pfotzer maximum during GLE $\#$ 65 on 28 October 2003.}. 
\label{Fig3}
\end{figure}

\begin{figure}
\centering
\includegraphics[width=0.8\textwidth]{{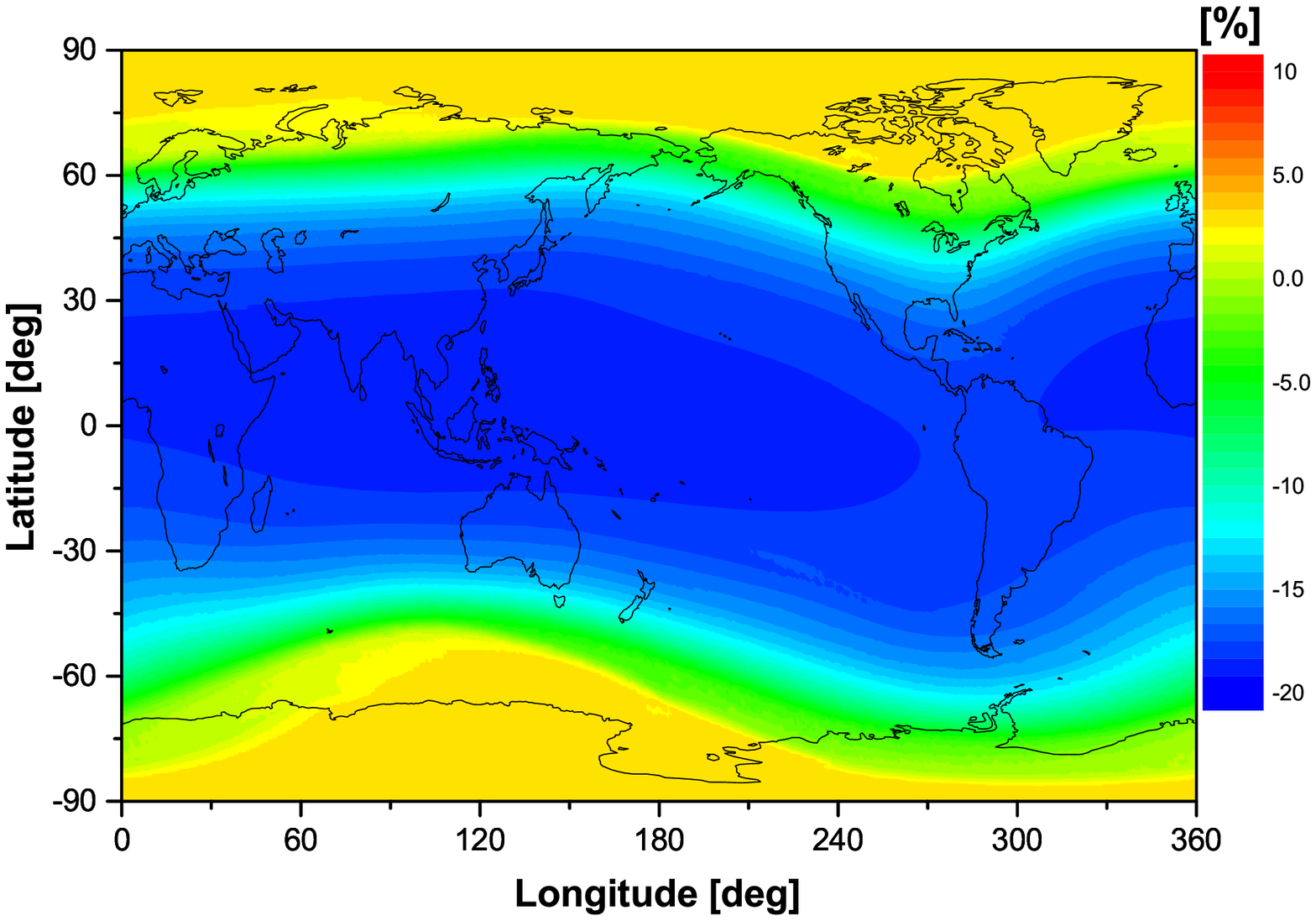}}
\caption{Map of the 24$^{h}$ averaged ionization effect in the region of Regener-Pfotzer maximum during GLE $\#$ 66 on 29 October 2003.}. 
\label{Fig4}
\end{figure}

\begin{figure}
\centering
\includegraphics[width=0.8\textwidth]{{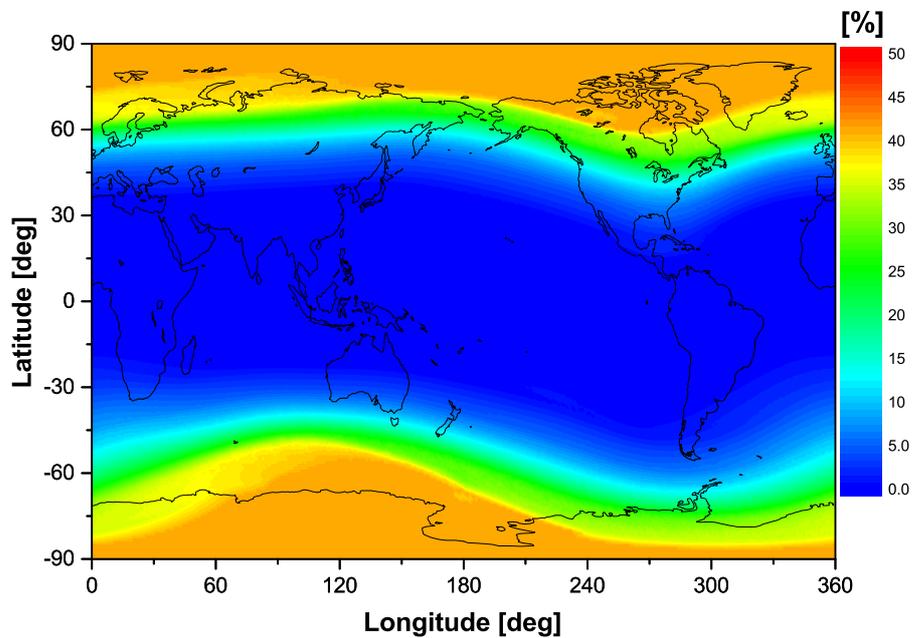}}
\caption{Map of the 24$^{h}$ averaged ionization effect in the region of Regener-Pfotzer maximum during GLE $\#$ 67 on 2 November 2003.}. 
\label{Fig5}
\end{figure}

The 24 hours integrated ionization effect during GLE $\#$ 65 was significant in the high-latitude region, where it ranged about 100-200 $\%$ in the region of Regener-Pfotzer maximum. The ionization effect diminished at lower latitudes, it was negligible in lower-mid and equatorial latitudes (Fig.3). This is due to softer SEP spectra comparing to GCRs. Therefore SEPs contributed significantly in the polar region, but not at low latitudes. The ionization effect is also a function of the altitude. It diminished significantly as a function of altitude. At lower altitudes, it was considerably smaller than that at the region of the Regener-Pfotzer maximum. An illustration of the altitude dependence of the event averaged ionization effect is given in Appendix (Fig.A.4). Note that the event integrated ionization effect was considerably greater than 24 hours integrated, since it accounted the SEPs contribution on a shorter time scale.

As was mentioned above, GLE $\#$ 66 occurred during a deep Forbush decrease of GCRs and complex magnetospheric conditions \citep[e.g.][]{Belov2005}. This specific Forbush decrease was one of the largest ever recorded. In particular, the amplitude of the decrease was $\sim$ 30$\%$ in 10 GV particles as reported by \citet{Belov2009} in their Figure 2. This resulted in complicated interplay of SEP contribution and significantly reduced GCR flux contribution on the ion production. The SEP spectra in this case were slightly softer and with reduced flux than that during GLE $\#$ 65.  While, the event integrated ionization effect was comparable to the previous event, the 24 hours averaged ionization effect during GLE $\#$ 66 was marginal, even in the polar region it was of about 5 $\%$. Moreover, it was negative in the equatorial region, due to the reduced GCR flux (Fig.4). The altitude dependence of the event  averaged ionization effect during  GLE $\#$ 66 is given in the Appendix (Fig.A.5). 

Accordingly, the ionization effect during GLE $\#$ 67 was not significant in mid-latitudes and it was marginal in the equatorial region (Fig.5). The last of the sequence of Halloween events, GLE $\#$ 67, occurred during the recovery phase of a deep Forbush decrease. Therefore, the reduced GCR flux within temporal evolution was explicitly considered for the computation of background ion rate production. The 24 hours averaged ionization effect during GLE $\#$ 67 was of about 40 $\%$ in the polar region. It diminished to about 10 $\%$ at lower latitudes. Accordingly, the altitude dependence of the event  averaged ionization effect during  GLE $\#$ 67 is given in the Appendix (Fig.A.6).

\section{Summary and Discussion}
The effect of high-energy particles precipitation, specifically cosmic rays, via the induced ionization on atmospheric chemistry and physics is subject to extensive scientific discussion \citep{Mironova2015}. The sporadic rapid change of CR flux, e.g. Forbush decreases and/or GLEs provides an unique possibility to study such possible effects on enhanced magnitude. The studied here sequence of three consecutive GLEs occurred in October-November 2003, the so-called Halloween events, give a good basis to study the possible influence of precipitating energetic particles, concerning also previous reports \citep[e.g.][]{Funke20119089}. 

It was recently pointed out that for observation of such effects, it is necessary an essential increase of ion production in the atmosphere e.g. during major GLEs and winter season in order to avoid the eventual influence of UV \citep{Mironova2014}. Therefore, the presented here computed ionization effect during Halloween events gives the basis for similar studies \citep[e.g.][]{Krivolutsky2005105, Sinnhuber20181115}. 

The ion production during the GLEs is governed by SEP spectra. SEPs with harder spectra impact also mid-latitude regions, while soft SEPs contribute mostly to the polar region. In addition, transients such as Forbushes play also an important role. This is clearly seen during the Halloween events. The ion production during GLE $\#$ 65 was greater than the subsequent GLE $\#$ 66 and GLE $\#$ 67. Accordingly, the ionization effect was highly dependent on the time-scale. While, the event averaged ionization effect was in the same order for all the three events, the 24 hours integrated ionization effect was apparently different: during GLE $\#$ 65 it was significant, marginal even negative in some regions during GLE $\#$ 66 and moderate during GLE $\#$ 67. Besides, an apparent altitude dependence was observed. 

In the work presented here, we computed the ion production rate and the corresponding ionization effect during the Halloween GLEs occurred on October-November 2003, assuming explicitly precisely the derived SEP spectra, their evolution throughout the events, and the corresponding magnetospheric conditions and variable GCRs flux. The computed ion production rates were significant during the main phase of the events, specifically at the polar region with rigidity cut-off $R_{c}$ $\le$ 1 GV. At regions with $R_{c}$ $\approx$ 2, the ion production was comparable to the average due to GCR, because of the rapidly falling SEP spectra. At mid-latitudes with rigidity cut-off $R_{c}$ of about 3 GV or greater, the ion production due to GCR dominated in the whole atmosphere during all the events. The event averaged ionization effects were maximal at altitudes of about 15--18 km a.s.l..   
  
The presented here computation of ionization effect during the sequence of three Halloween events give good basis to study the possible effect of precipitating high-energy particles in the Earth's atmosphere on a minor constituents, atmospheric physics and chemistry as well as studies related to space weather and solar-terrestrial physics \citep{Miroshnichenko03, Mishev2015359, Miroshnichenko18}.

\section*{Acknowledgements}
This work was supported by the Academy of Finland (project 330063 QUASARE, 321882 ESPERA and  304435 CRIPA-X). The work benefited from discussions in the framework of the International Space Science Institute International Team 441: High EneRgy sOlar partICle Events Analysis (HEROIC). The NM records used in this study were retrieved from International GLE database \texttt{http://gle.oulu.fi} and neutron monitor database \texttt{www.NMDB.eu}. The authors acknowledge the anonymous reviewers for their useful comments and suggestions that helped us to improve this paper.

\newpage
\begin{appendix} 
\section{Time and altitude evolution of ion production rate during the Halloween GLEs}
\setcounter{figure}{0}

\counterwithin{figure}{section}

\begin{figure}[htp]
   \centering
   \includegraphics[width=0.3\textwidth]{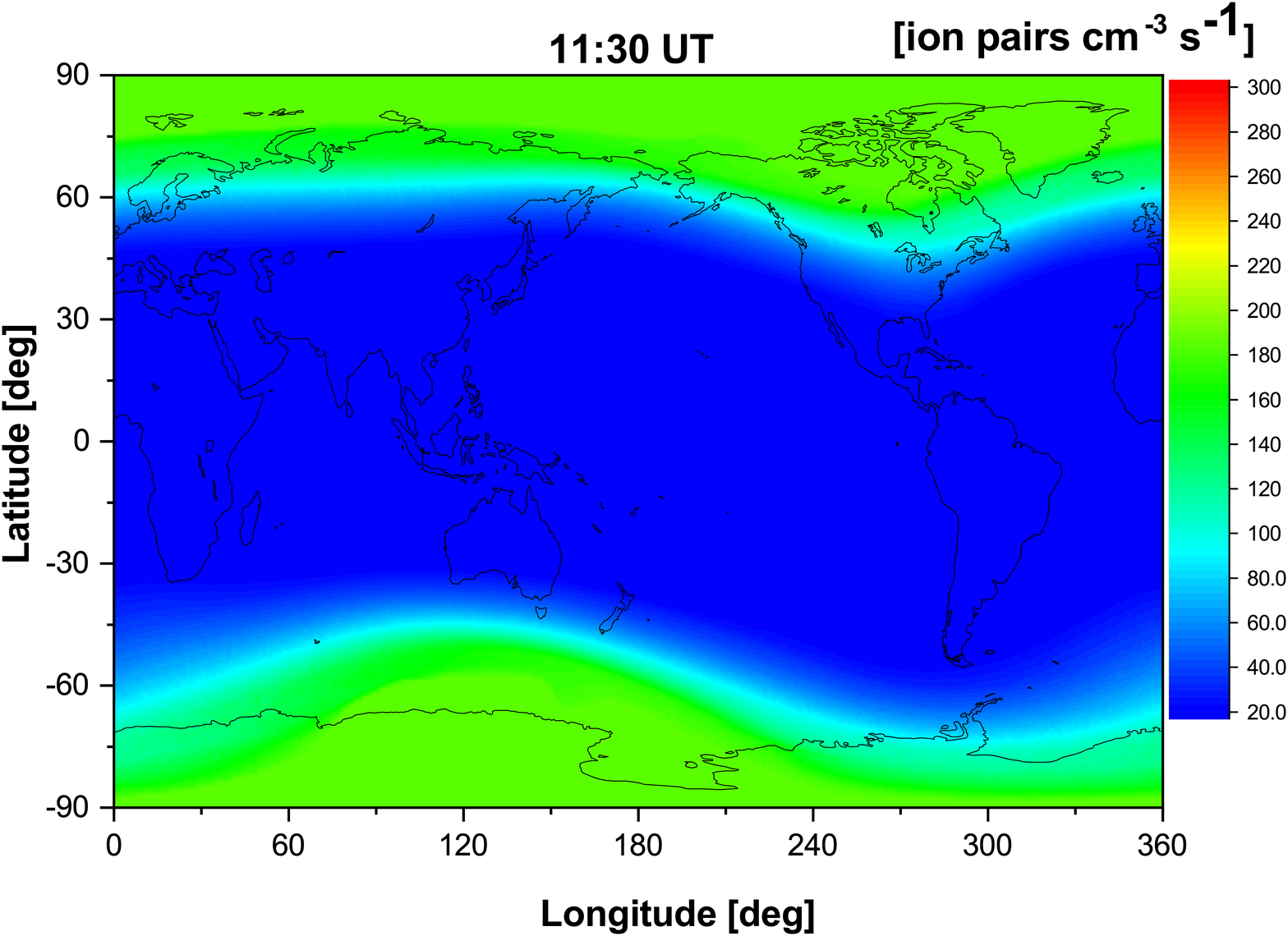}\quad
    \includegraphics[width=0.3\textwidth]{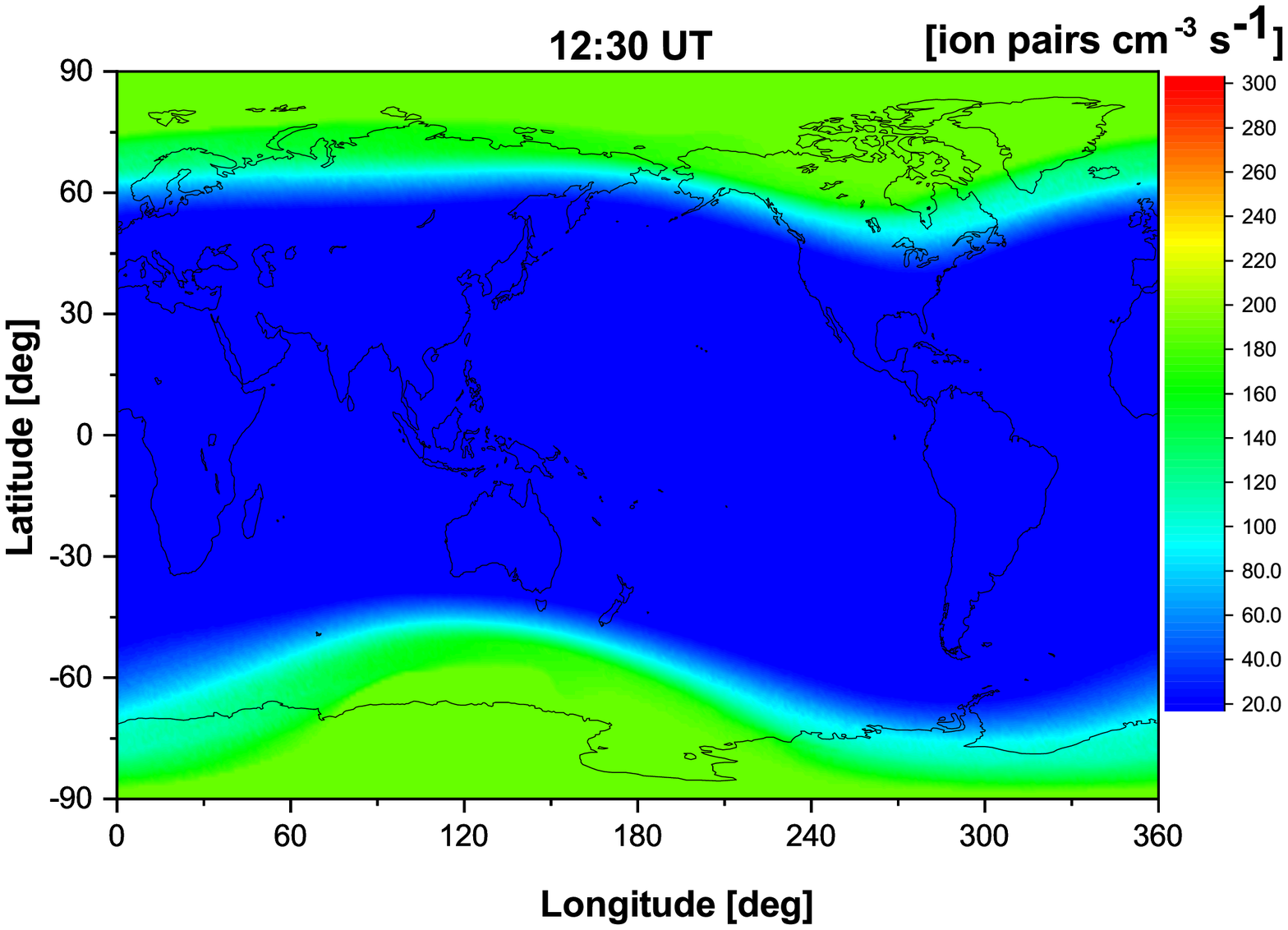}
    
    \medskip
   \includegraphics[width=0.3\textwidth]{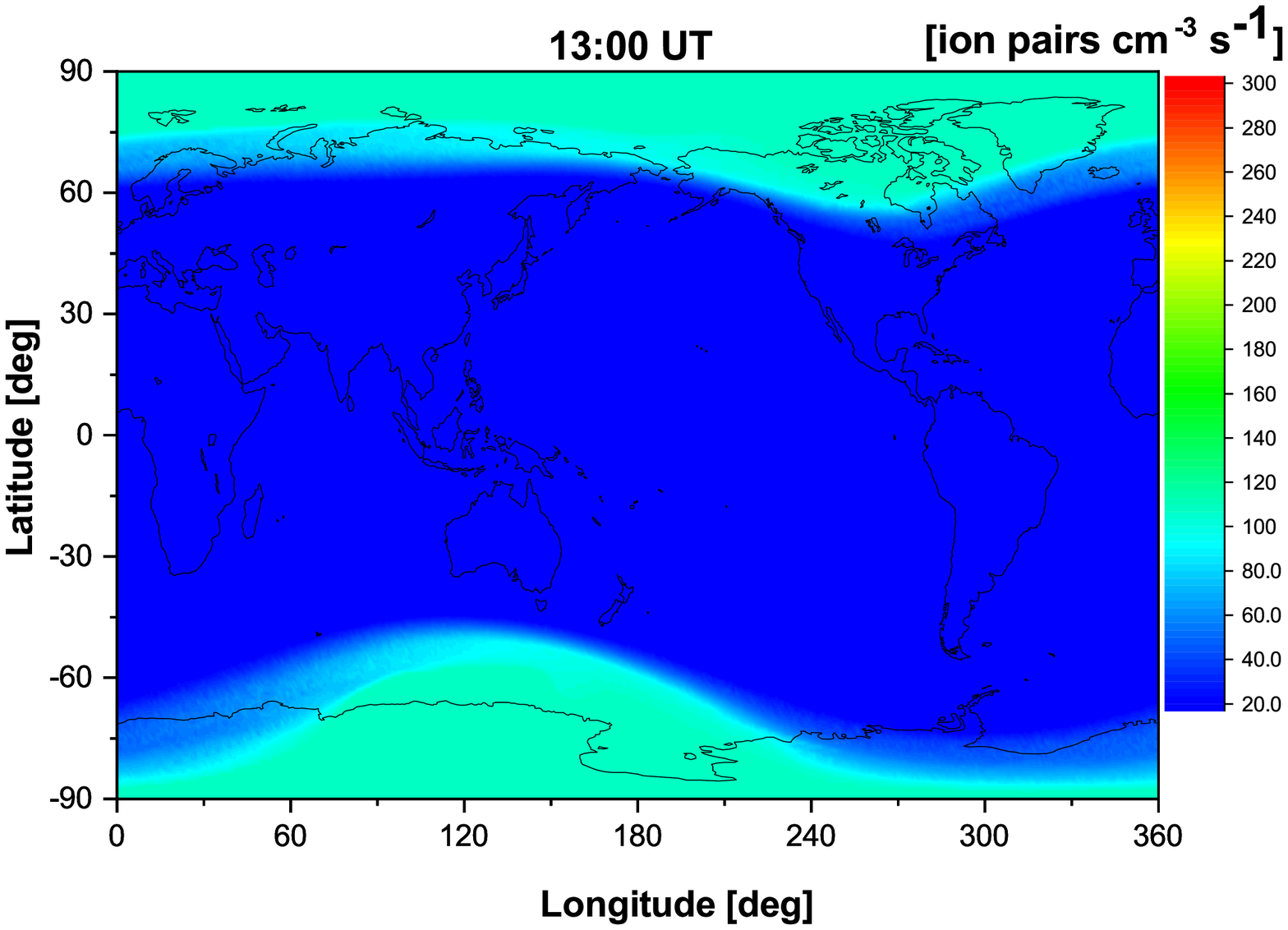}\quad
    \includegraphics[width=0.3\textwidth]{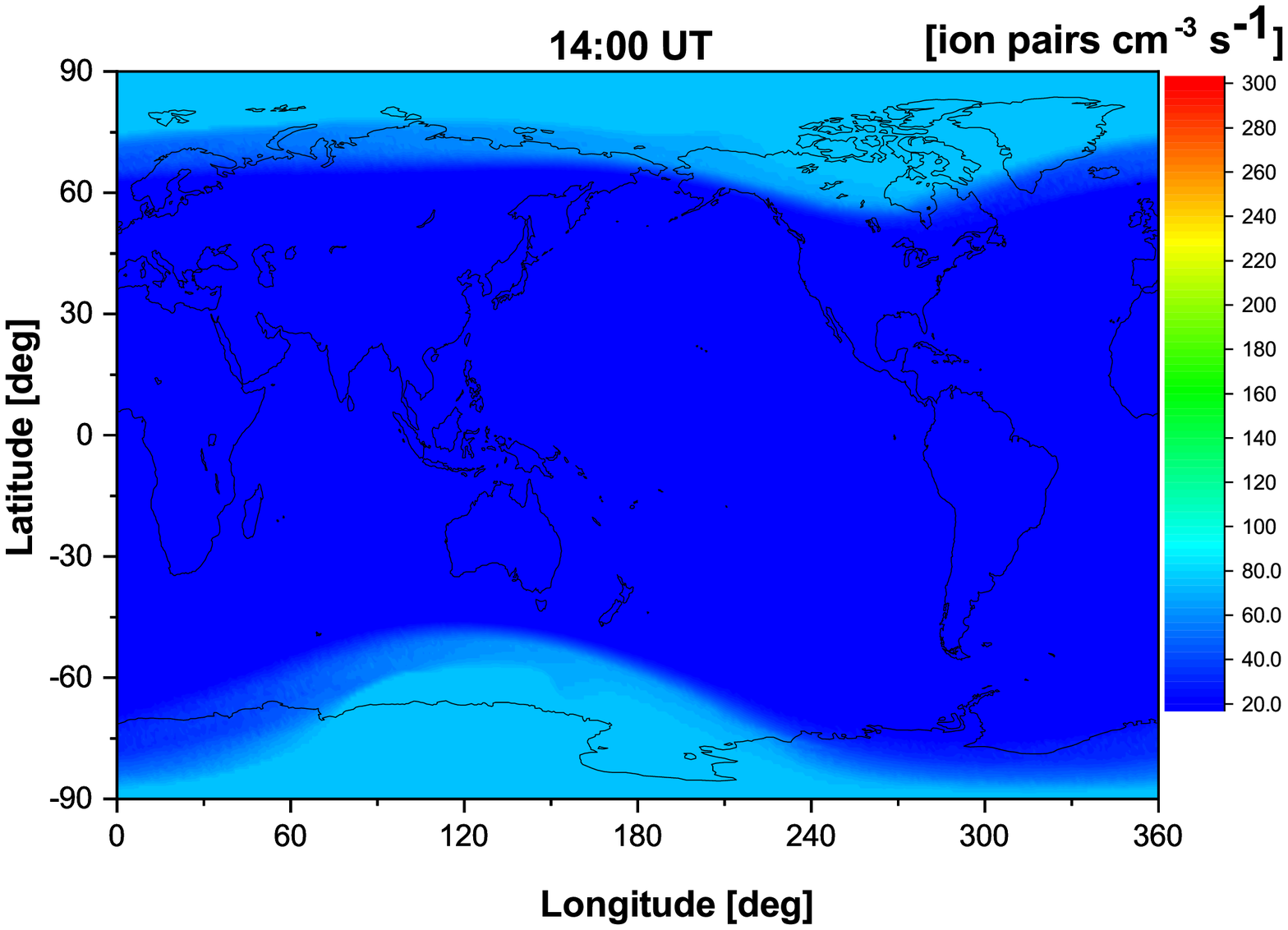}

      \caption{Ion production rate at altitude of 15 km a.s.l. at selected stages of the event during  GLE $\#$ 65 on 28 October 2003.
              }
         \label{FigA1_1}
   \end{figure}

\begin{figure}[htp]
   \centering
   \includegraphics[width=0.3\textwidth]{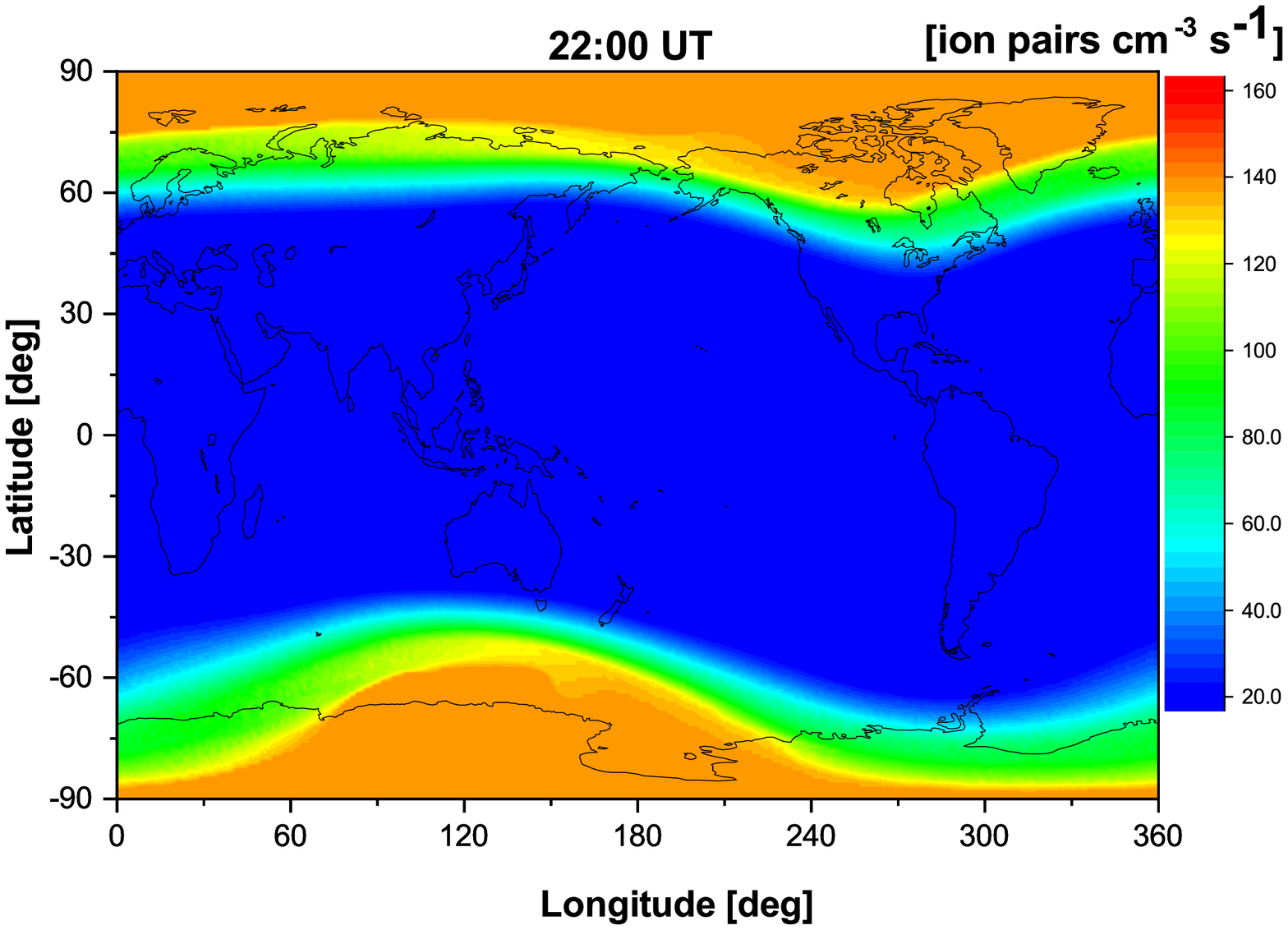}\quad
    \includegraphics[width=0.3\textwidth]{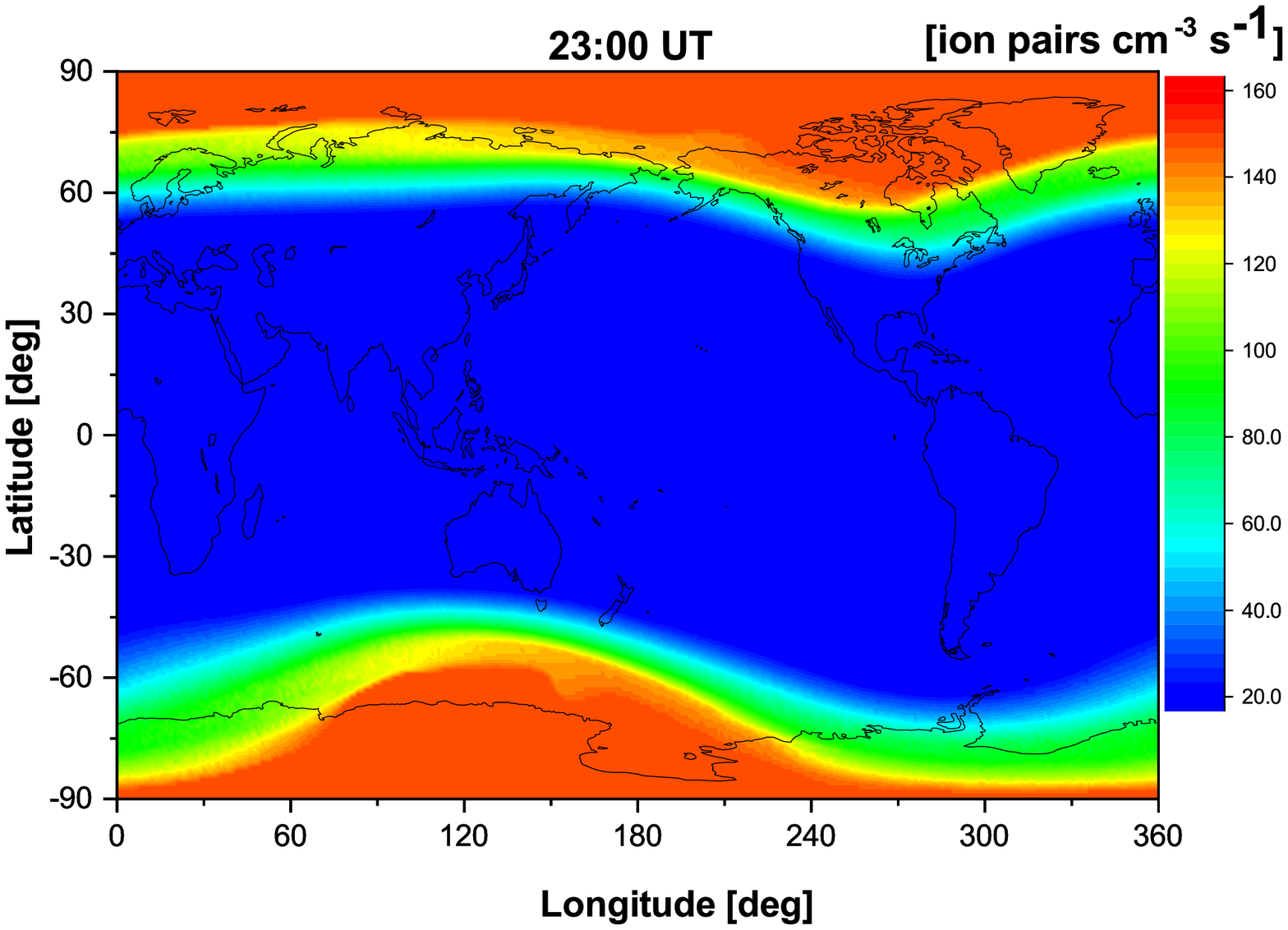}
    
    \medskip
   \includegraphics[width=0.3\textwidth]{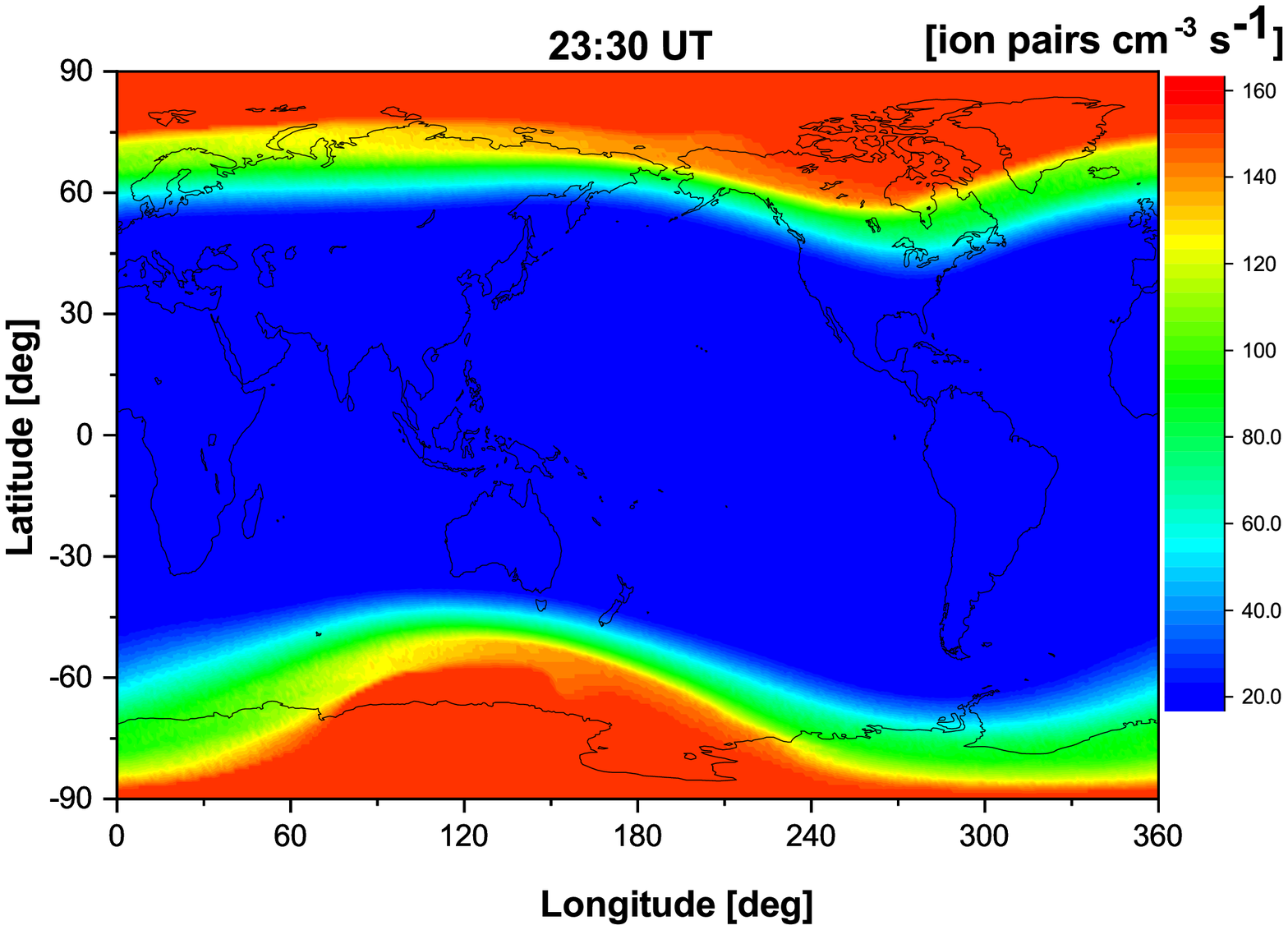}\quad
    \includegraphics[width=0.3\textwidth]{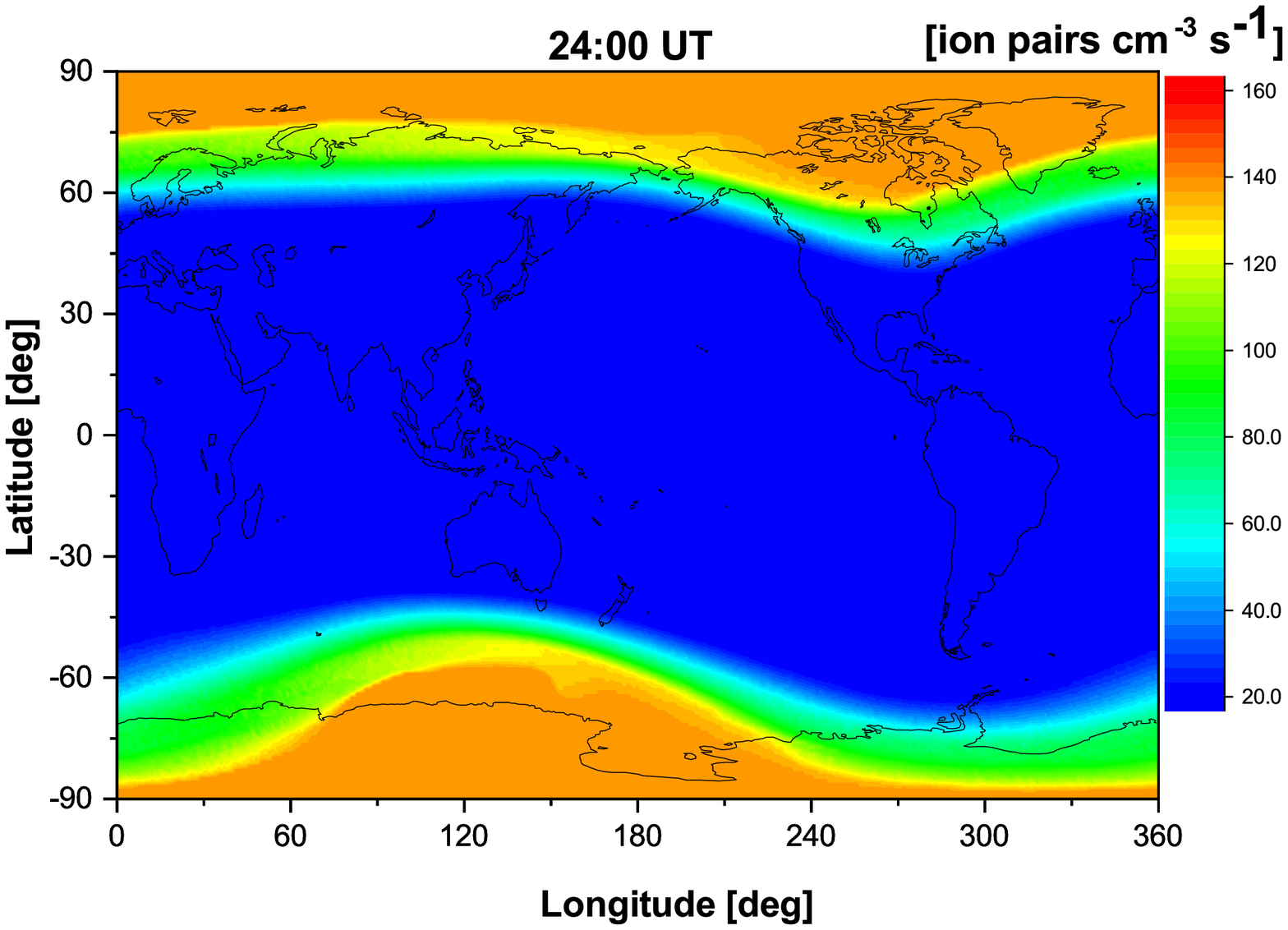}

      \caption{Ion production rate at altitude of 15 km a.s.l. at selected stages of the event during  GLE $\#$ 66 on 29 October 2003.
              }
         \label{FigA1_2}
   \end{figure}

\begin{figure}[htp]
   \centering
   \includegraphics[width=0.3\textwidth]{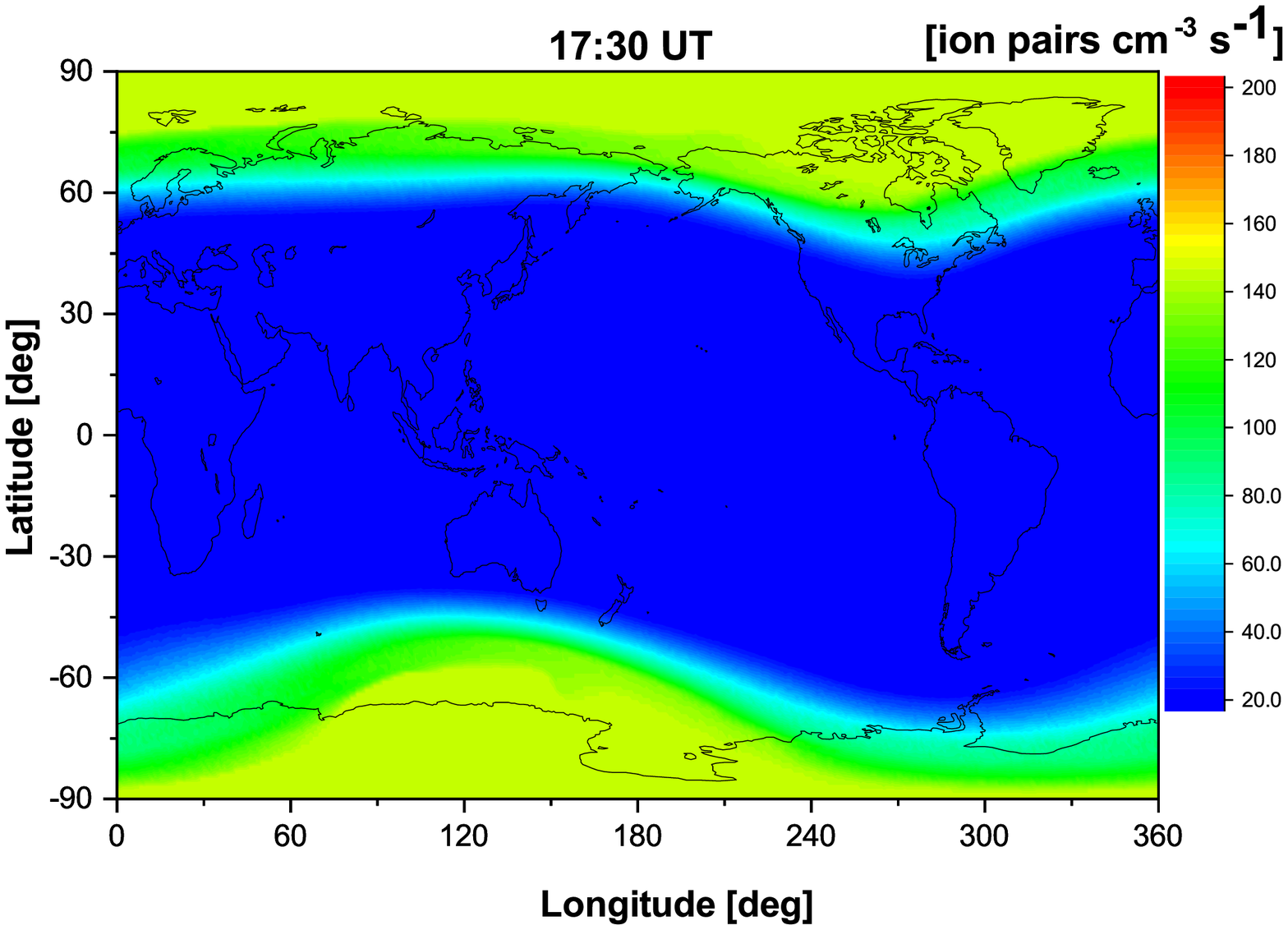}\quad
    \includegraphics[width=0.3\textwidth]{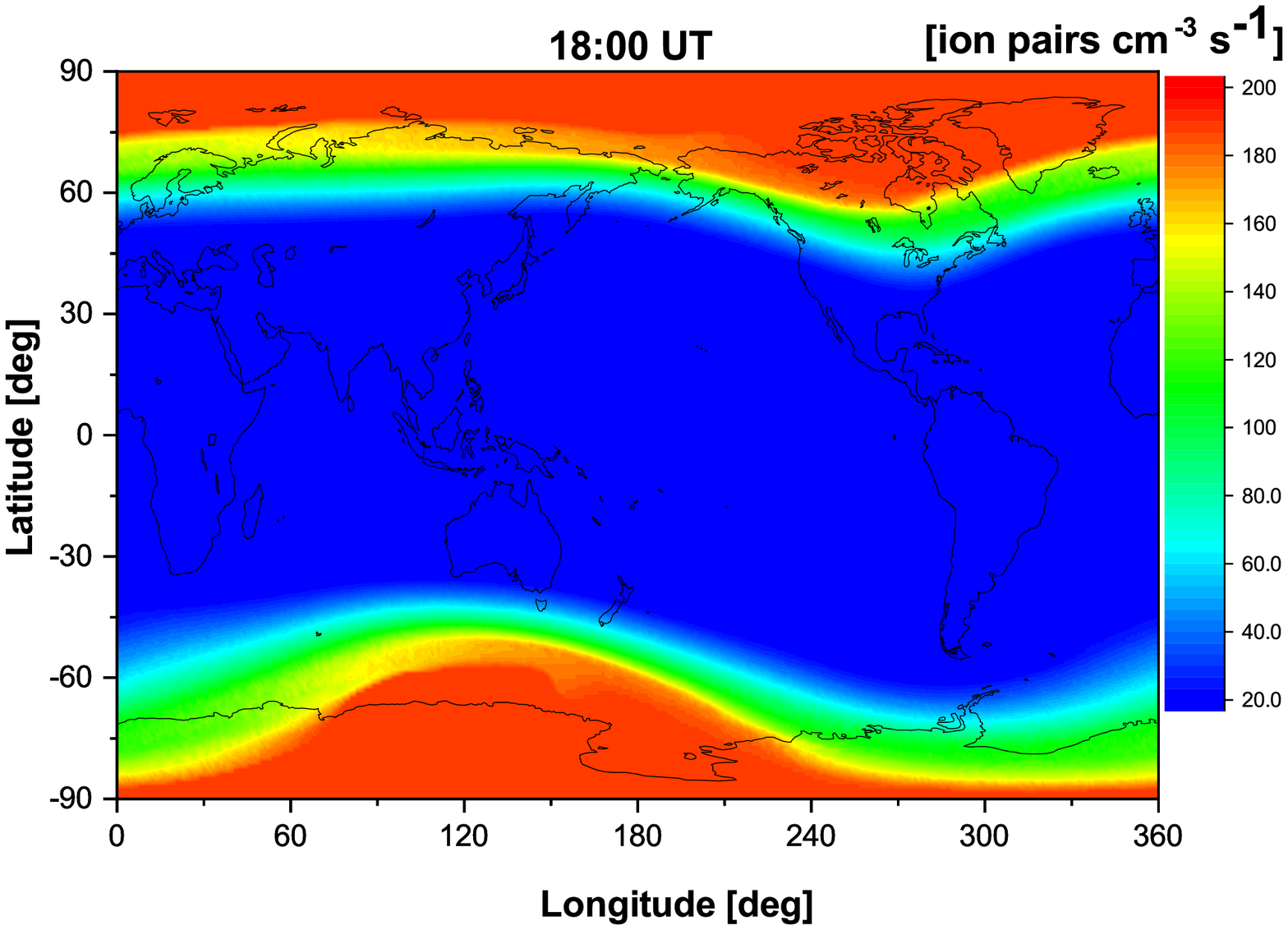}
    
    \medskip
   \includegraphics[width=0.3\textwidth]{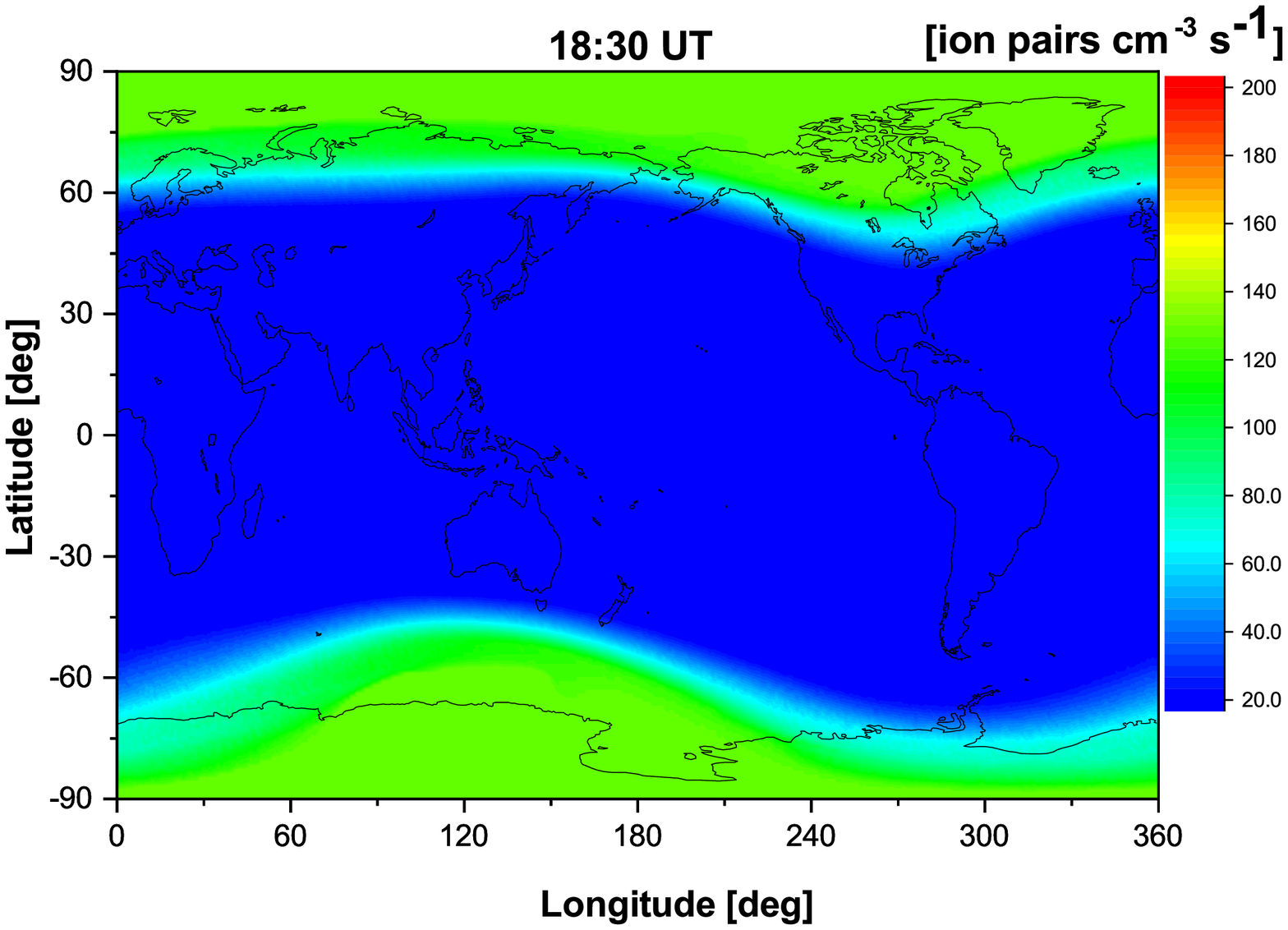}\quad
    \includegraphics[width=0.3\textwidth]{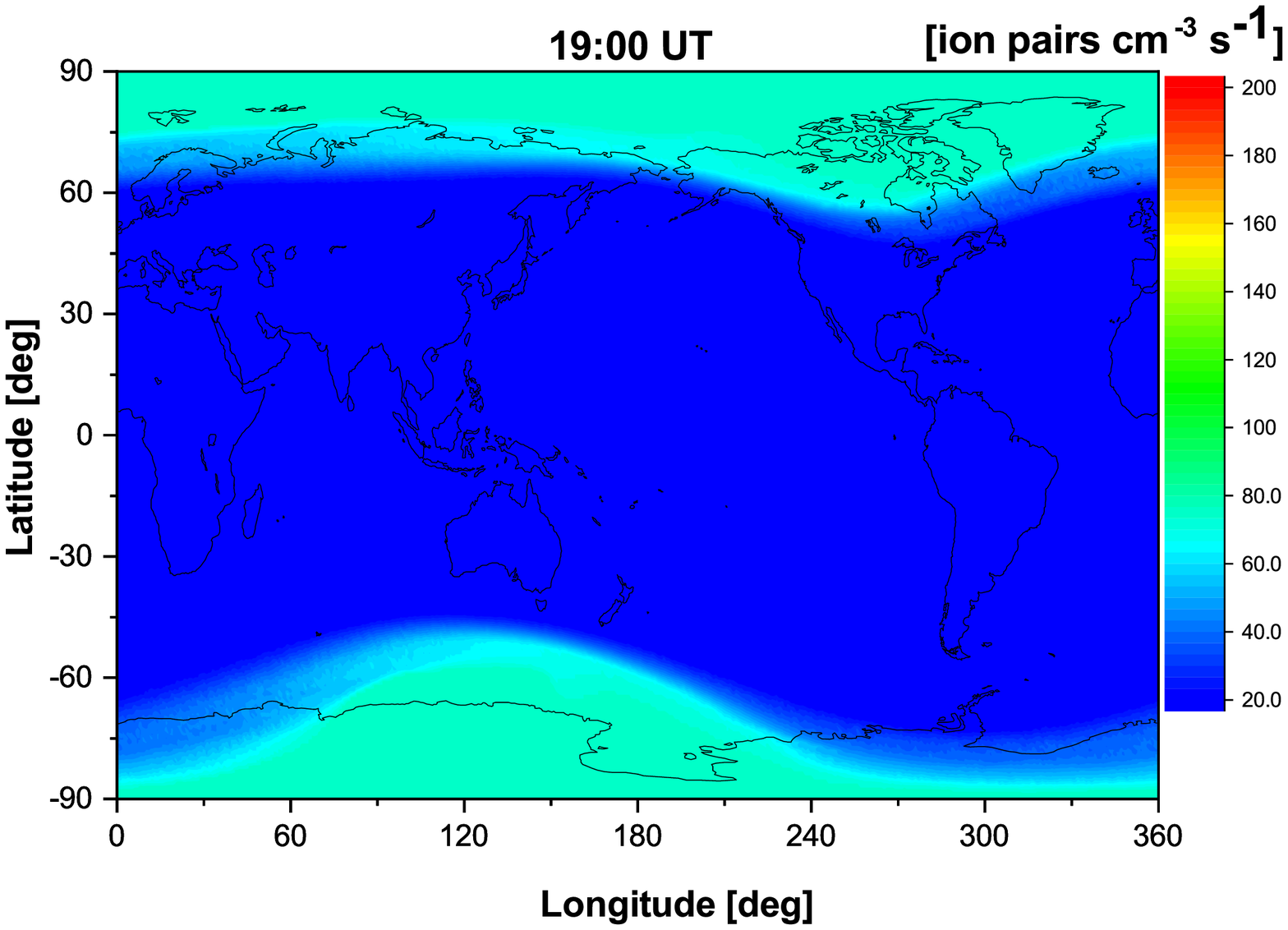}

      \caption{Ion production rate at altitude of 15 km a.s.l. at selected stages of the event during  GLE $\#$ 67 on 2 November 2003.
              }
         \label{FigA1_3}
   \end{figure}

\begin{figure}[htp]
   \centering
   \includegraphics[width=0.3\textwidth]{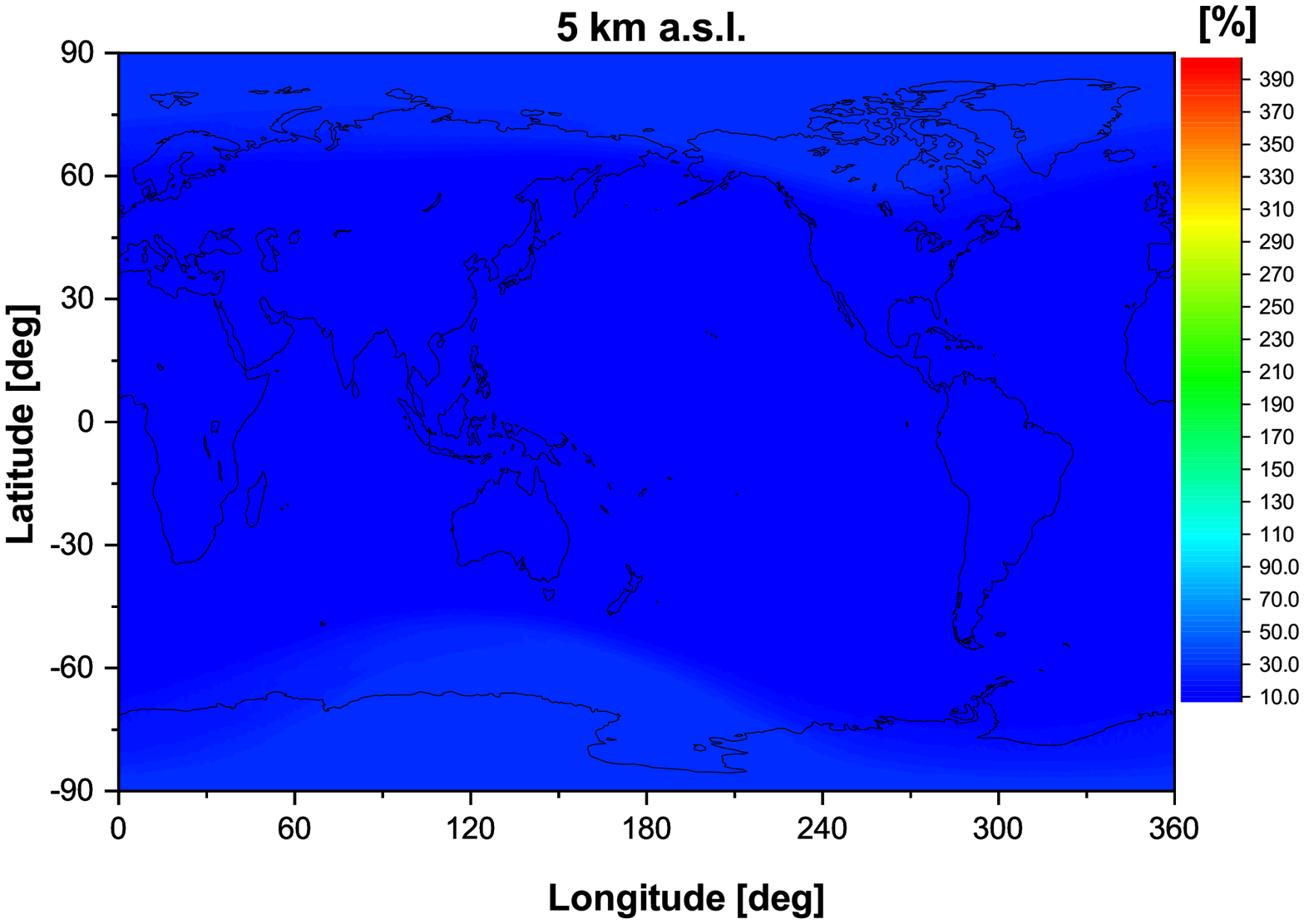}\quad
    \includegraphics[width=0.3\textwidth]{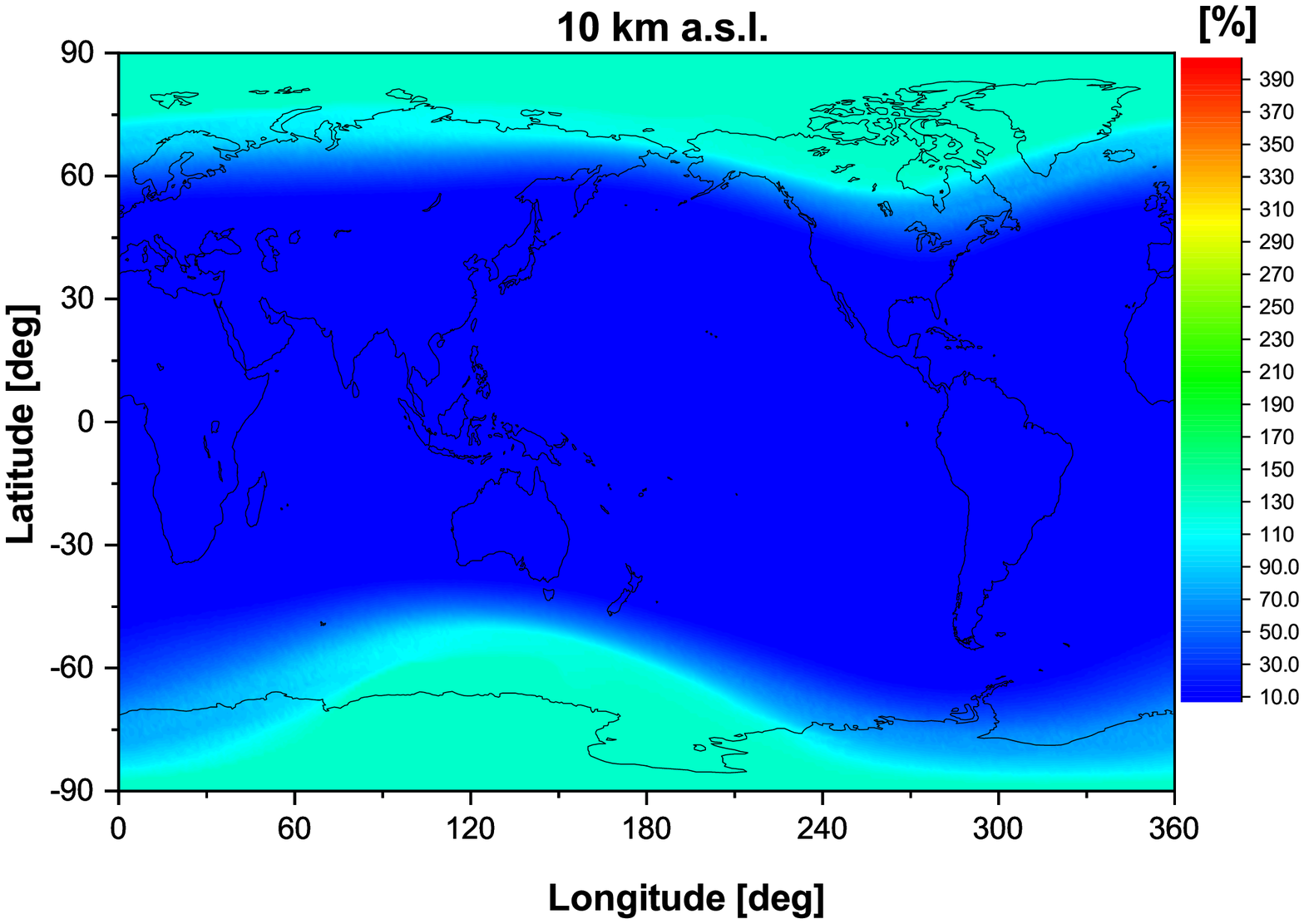}

    \medskip
   \includegraphics[width=0.3\textwidth]{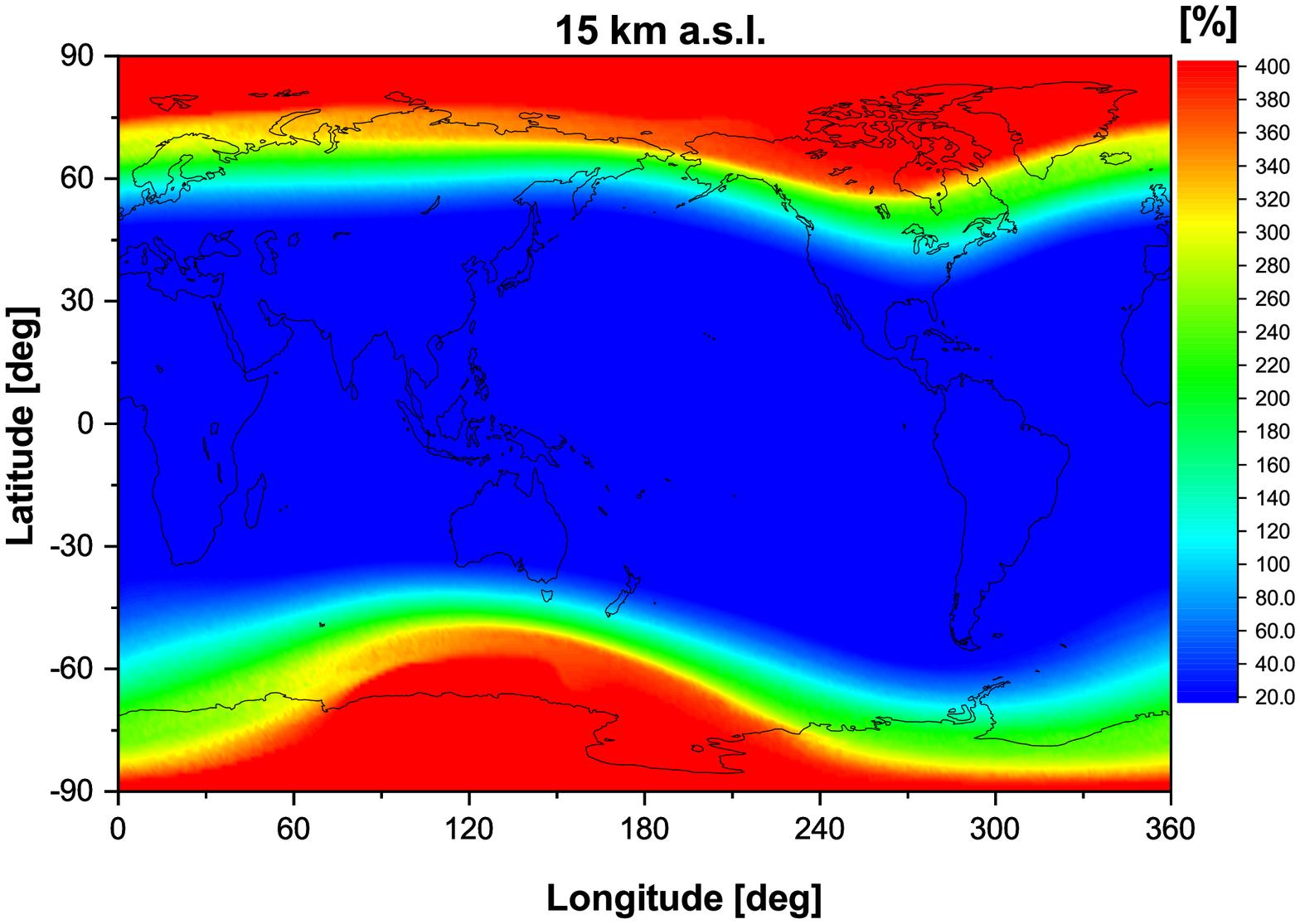}\quad
    \includegraphics[width=0.3\textwidth]{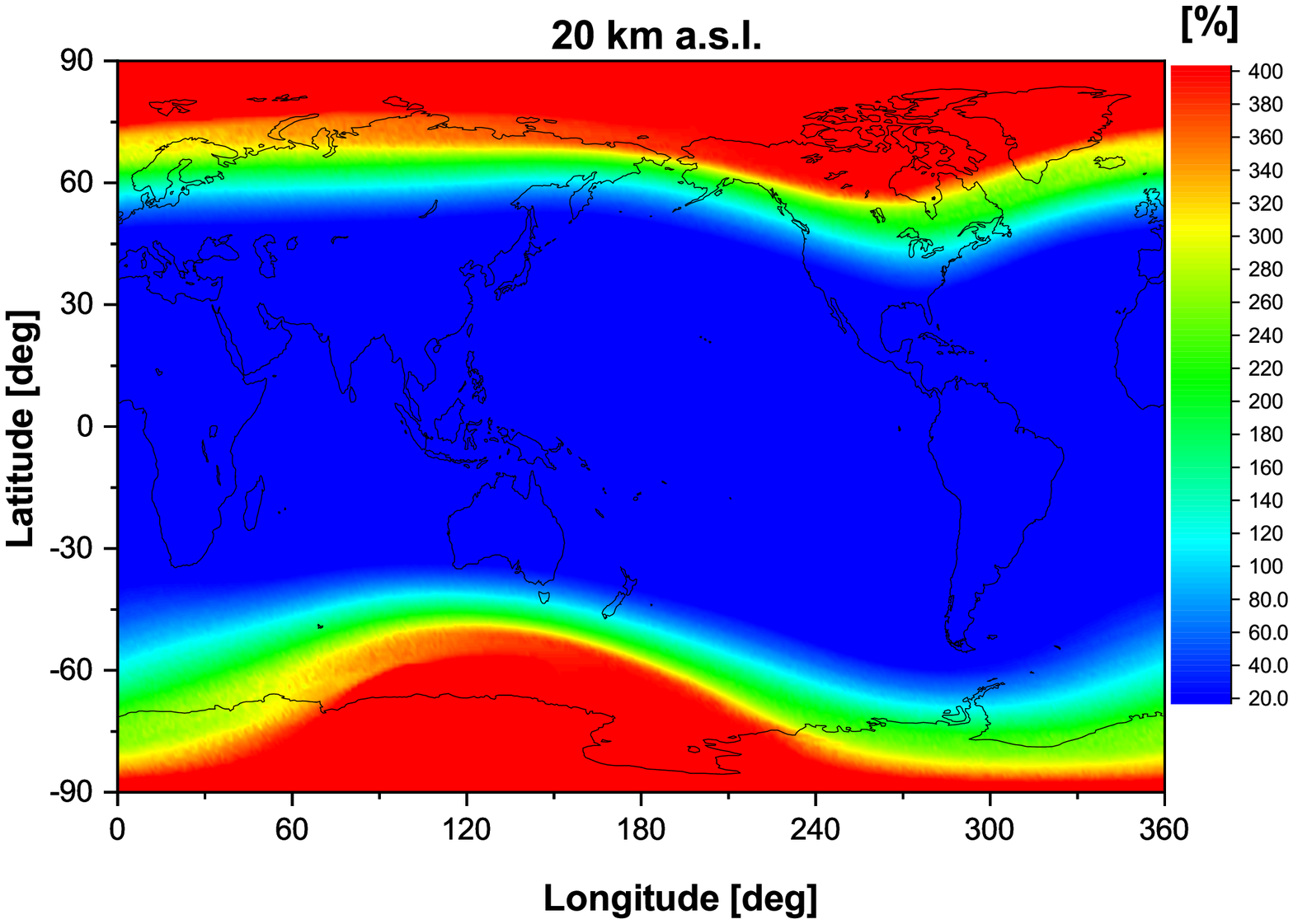}

      \caption{Event integrated ionization effect as a function of altitude during  GLE $\#$ 65 on 28 October 2003.
              }
         \label{FigA1_4}
   \end{figure}
   
   \begin{figure}[htp]
   \centering
   \includegraphics[width=0.3\textwidth]{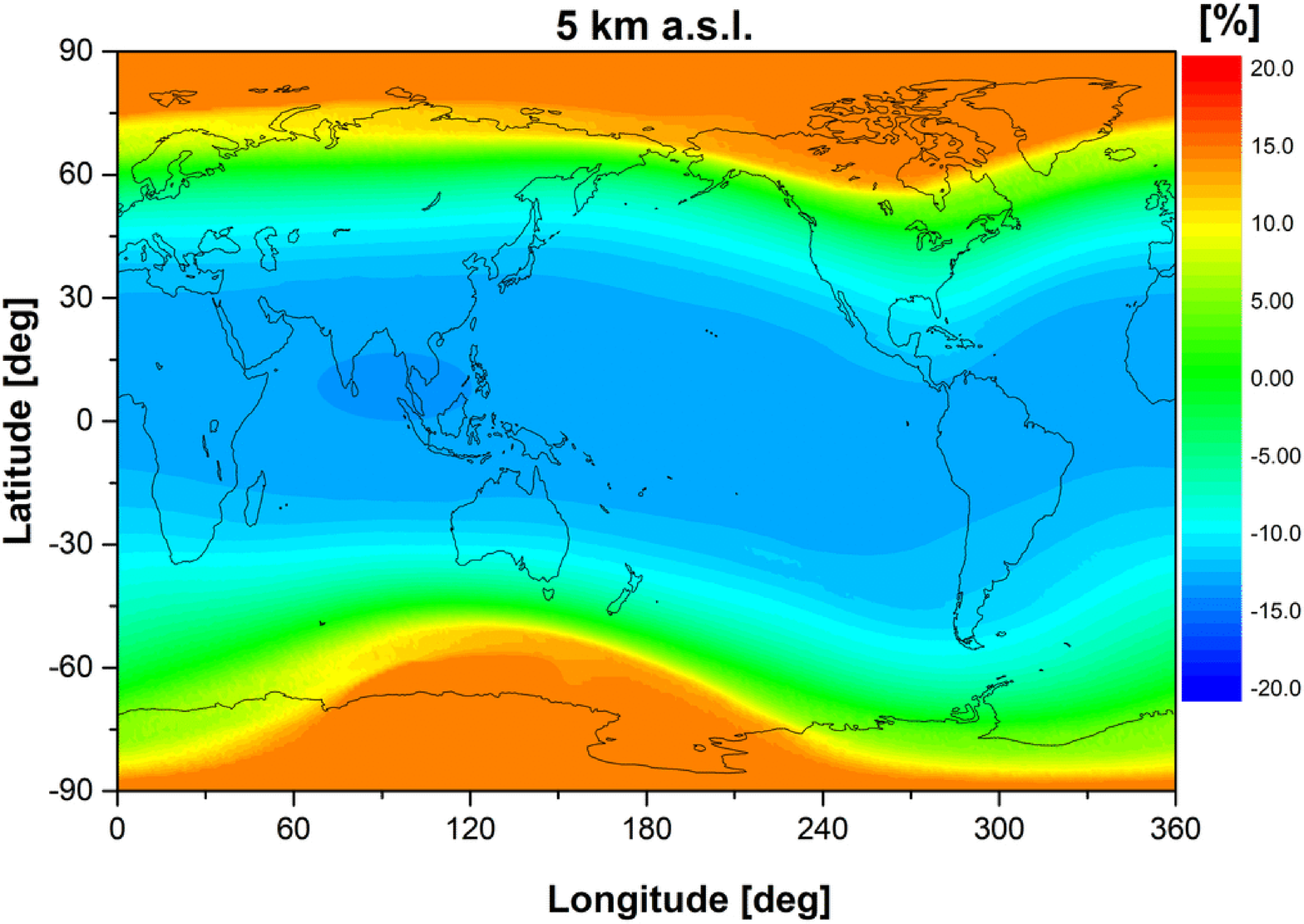}\quad
    \includegraphics[width=0.3\textwidth]{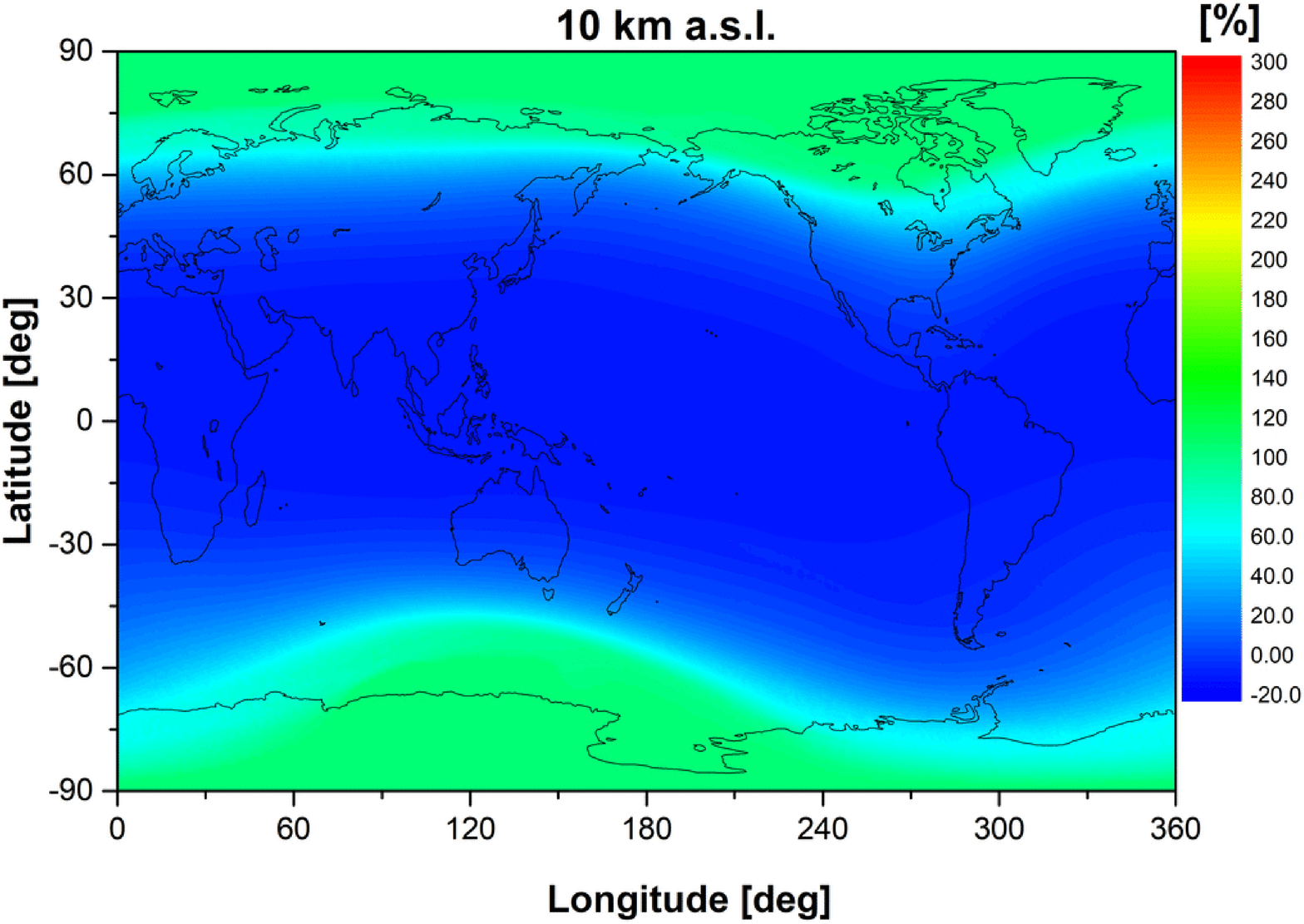}

    \medskip
   \includegraphics[width=0.3\textwidth]{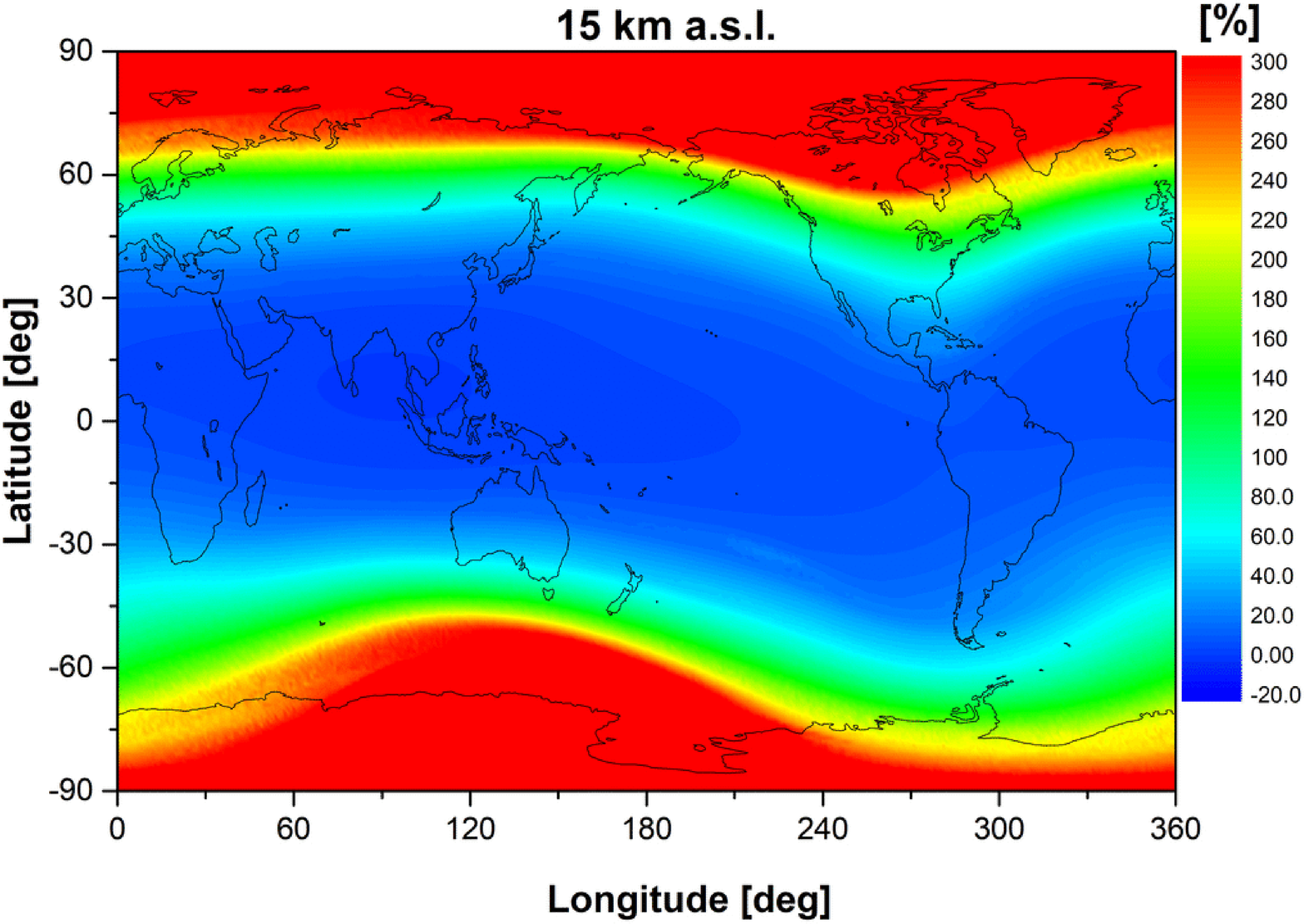}\quad
    \includegraphics[width=0.3\textwidth]{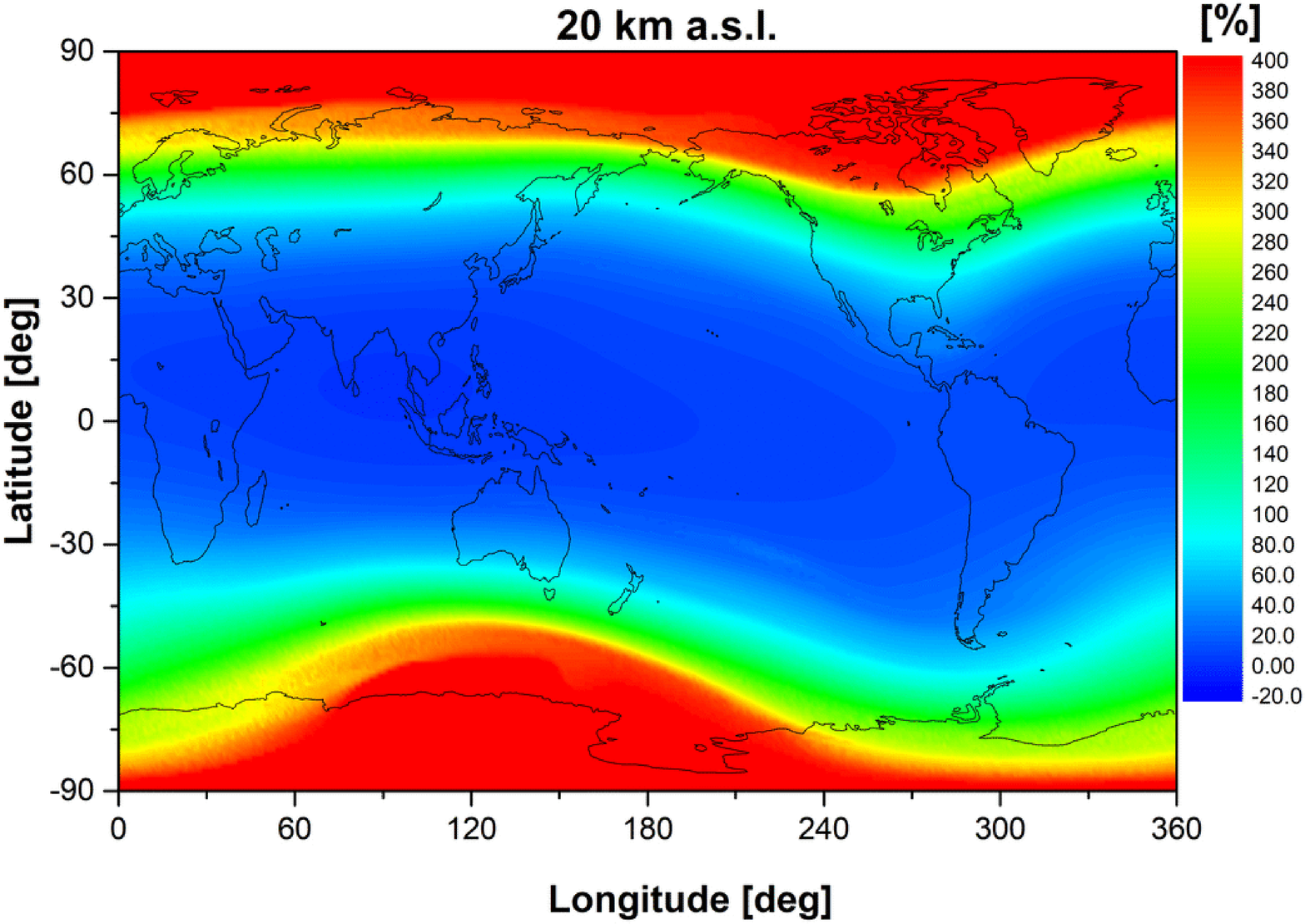}

      \caption{Event integrated ionization effect as a function of altitude during  GLE $\#$ 66 on 29 October 2003.
              }
         \label{FigA1_5}
   \end{figure}

 \begin{figure}[htp]
   \centering
   \includegraphics[width=0.3\textwidth]{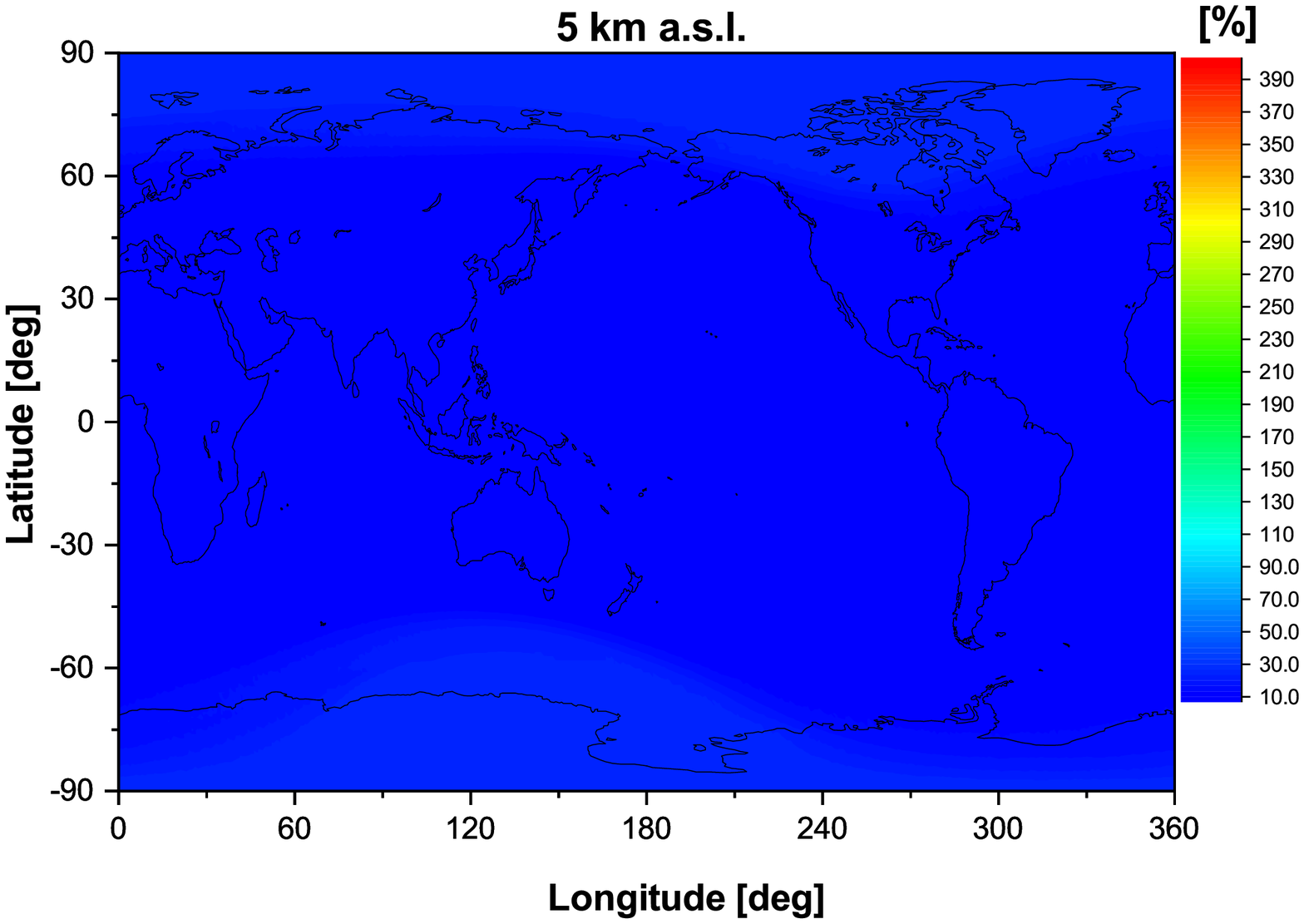}\quad
    \includegraphics[width=0.3\textwidth]{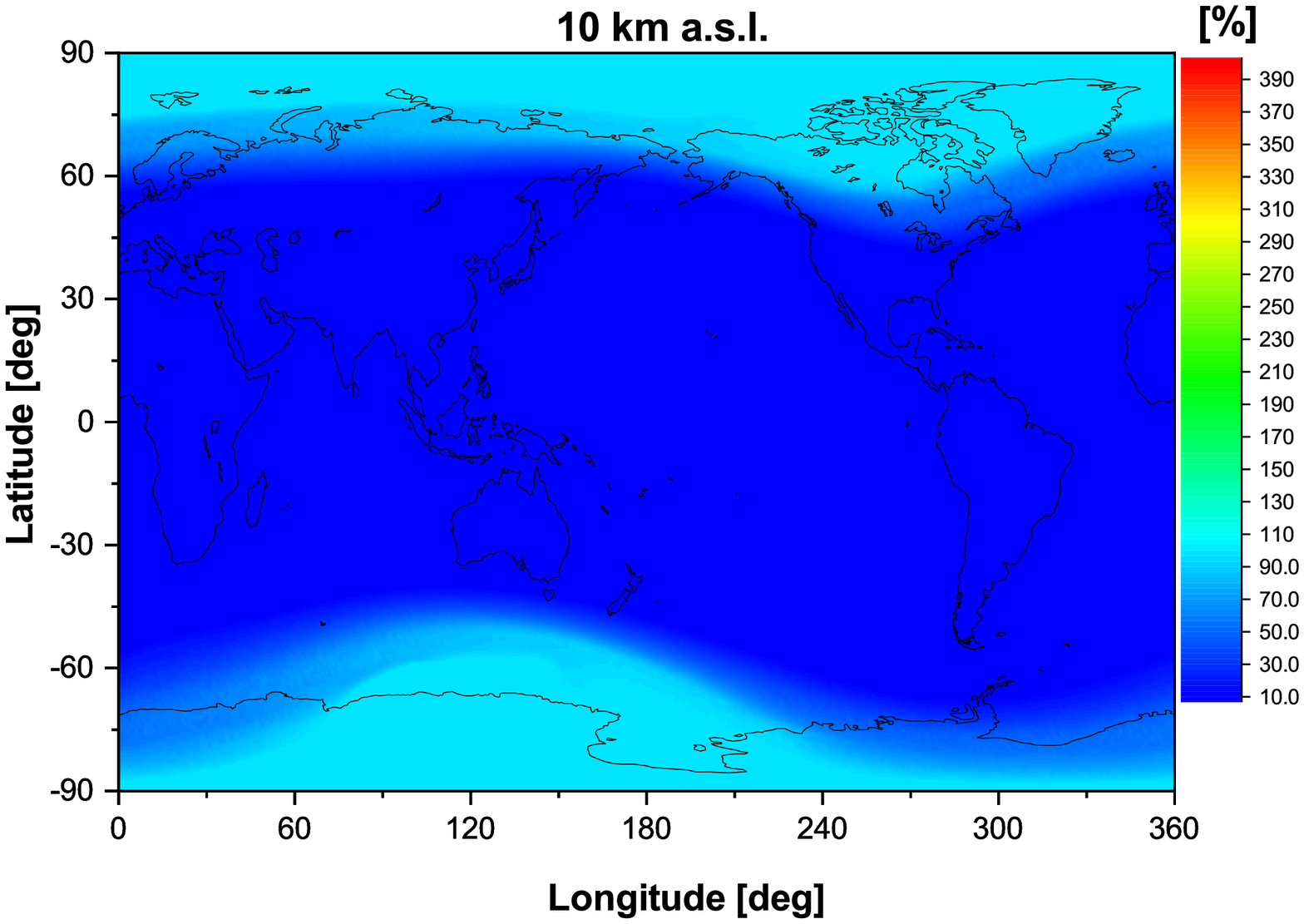}

    \medskip
   \includegraphics[width=0.3\textwidth]{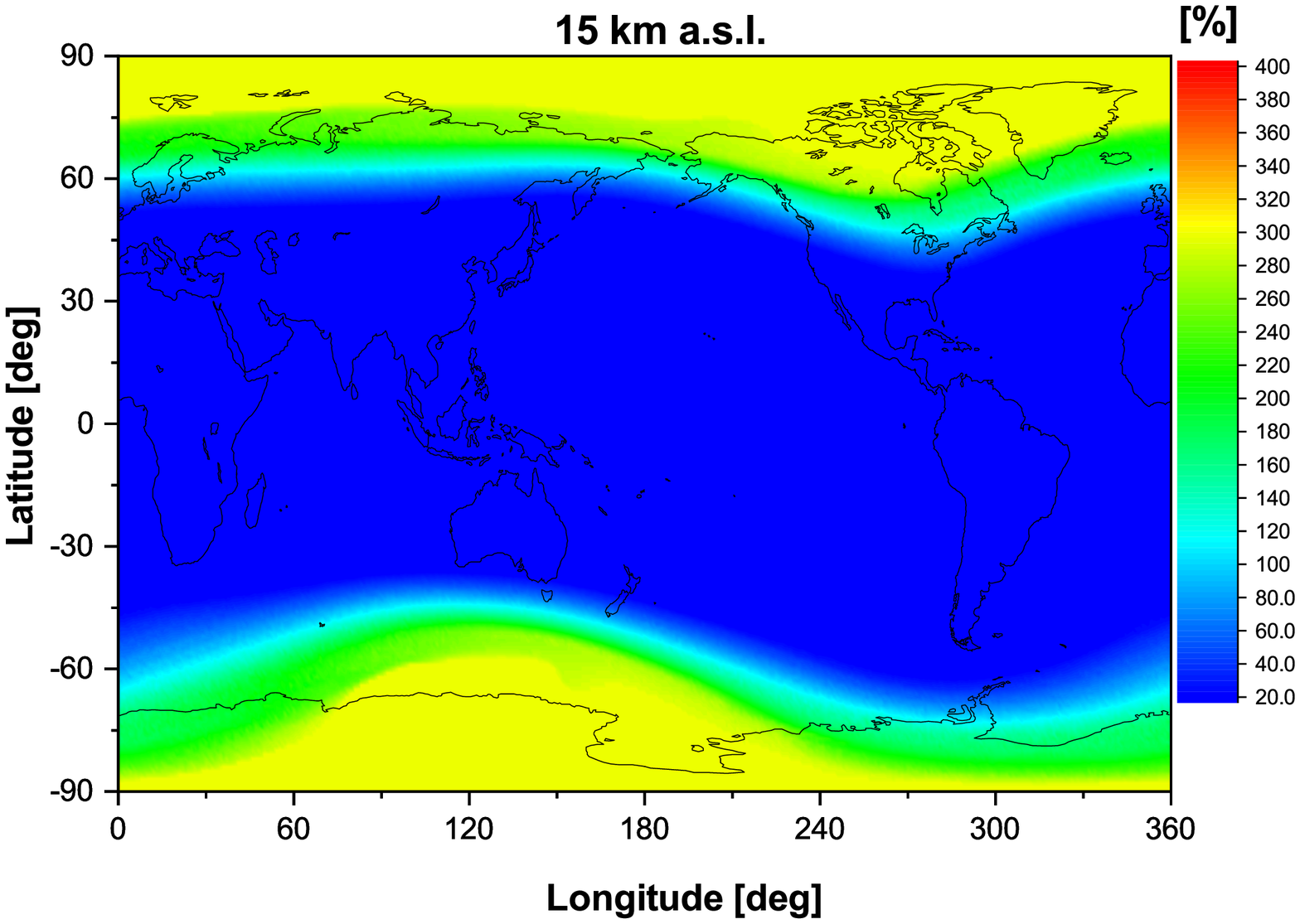}\quad
    \includegraphics[width=0.3\textwidth]{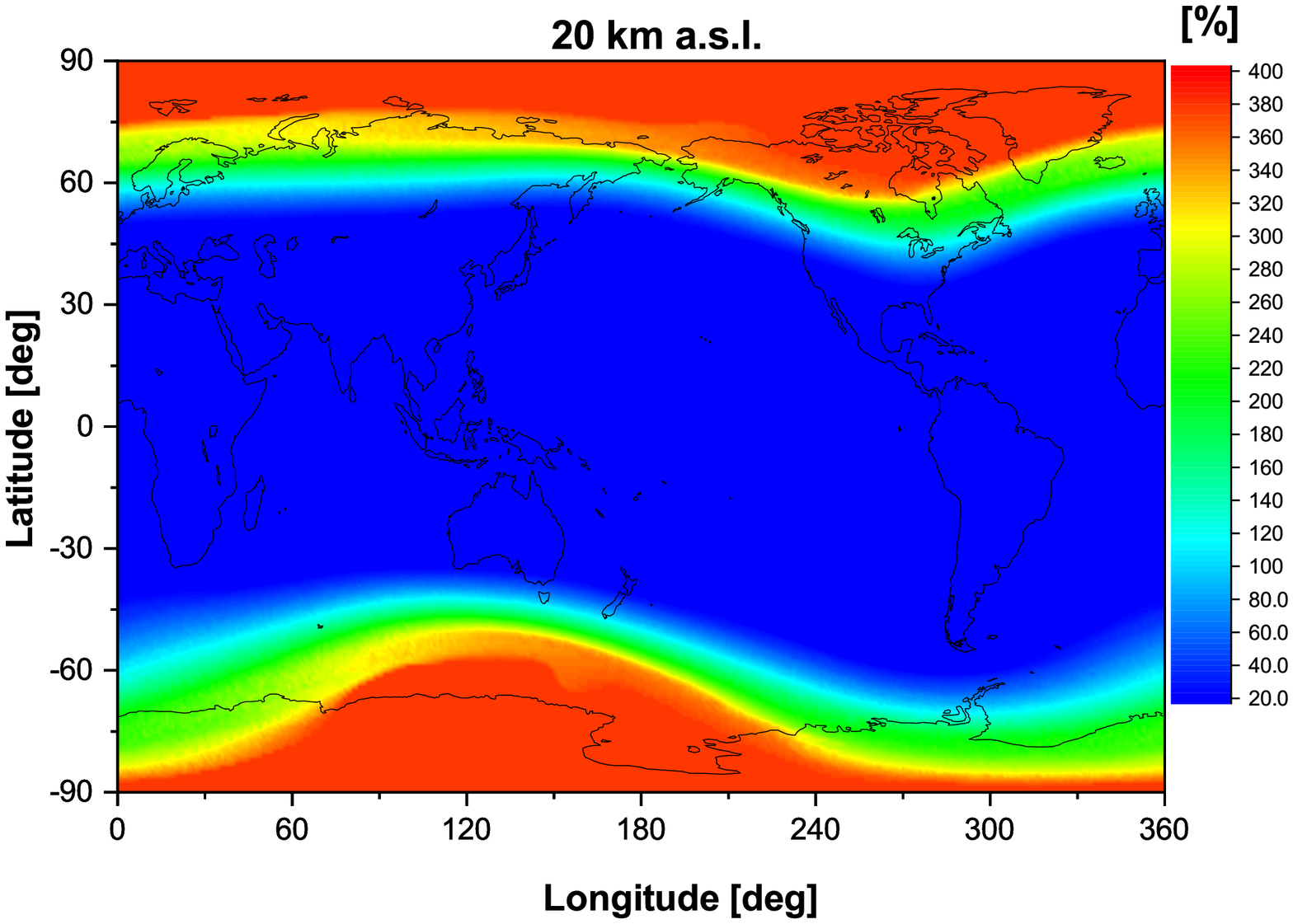}

      \caption{Event integrated ionization effect as a function of altitude during  GLE $\#$ 67 on 2 November 2003.
              }
         \label{FigA1_6}
   \end{figure}

\end{appendix}



\end{document}